\documentclass[aps,pra,twocolumn,10pt,notitlepage,showpacs]{revtex4-1}

\usepackage{color}
\usepackage[dvipsnames]{xcolor}
\usepackage{tikz}
\usetikzlibrary{calc,patterns}
\usepackage{ifthen}
\usepackage{amssymb}
\usepackage{mathtools}
\usepackage{dsfont}
\usepackage{bbm}
\usepackage{mleftright}\mleftright
\usepackage{thmtools} 
\usepackage{thm-restate}

\usepackage{algorithm2e}
\usepackage{enumitem}
\usepackage{float}
\usepackage{booktabs}
\usepackage{multirow}
\usepackage{color}

\usepackage{caption}
\usepackage{subcaption}

\newcommand{\nrm}[1]{\left\lVert #1 \right\rVert}

\newcommand{\pvp}{\vec{p}{\kern 0.45mm}'}
\let\oldnabla\nabla
\renewcommand{\nabla}{\oldnabla\!}

\DeclarePairedDelimiter\bra{\langle}{\rvert}
\DeclarePairedDelimiter\ket{\lvert}{\rangle}
\DeclarePairedDelimiterX\braket[2]{\langle}{\rangle}{#1 \delimsize\vert #2}
\newcommand{\underflow}[2]{\underset{\kern-60mm \overbrace{#1} \kern-60mm}{#2}}

\long\def\ignore#1{}

\newtheorem{theorem}{Theorem}

\newtheorem{lemma}[theorem]{Lemma}

\newcommand{\De}{\ensuremath{\mathcal{D}}}

\newcommand{\G}{\ensuremath{\mathcal{G}}}

\newcommand{\C}{\ensuremath{\mathcal{C}}}

\newcommand{\Oo}{\ensuremath{\mathcal{O}}}

\newcommand{\M}{\mathcal{M}}
\newcommand{\CG}{\ensuremath{\mathcal{C G}}}
\usepackage{tikz}

\usetikzlibrary{backgrounds,fit,decorations.pathreplacing,calc}

\usepackage{qcircuit}

\newenvironment{proof}
{\noindent {\bf Proof. }}
{{\hfill $\Box$}\\	\smallskip}

\usepackage[final]{hyperref}
\hypersetup{
	colorlinks = true,
	allcolors = {blue},
}

\usepackage{comment}
\usepackage{amsmath}

\usepackage{graphicx}
\usepackage{subcaption}
\usepackage{caption}
\usepackage{amssymb}
\usepackage{amsfonts}
\usepackage[parfill]{parskip}
\usepackage{hyperref}

\usepackage[format=plain,justification=centerlast,singlelinecheck=true]{caption}

\begin{document}


\title{On the optimality of spatial search by continuous-time quantum walk}


\author{Shantanav Chakraborty}  
\email[]{shchakra@ulb.ac.be}
\author{Leonardo Novo}
\email[]{lfgnovo@gmail.com}
\author{J\'{e}r\'{e}mie Roland}
\email[]{jroland@ulb.ac.be}
\affiliation{QuIC, Ecole Polytechnique de Bruxelles, Universit\'{e} Libre de Bruxelles}
\begin{abstract}
One of the most important algorithmic applications of quantum walks is to solve spatial search problems. A widely used quantum algorithm for this problem, introduced by Childs and Goldstone [Phys. Rev. A 70, 022314 (2004)], finds a marked node on a graph of $n$ nodes via a continuous-time quantum walk. This algorithm is said to be optimal if it can find any of the nodes in $O(\sqrt{n})$ time. However, given a graph, no general conditions for the optimality of the algorithm are known and previous works demonstrating optimal quantum search for certain graphs required an instance-specific analysis. In fact, the demonstration of necessary and sufficient conditions a graph must fulfill for quantum search to be optimal has been a long-standing open problem.

In this work, we make significant progress towards solving this problem. We derive general expressions, depending on the  spectral properties of the Hamiltonian driving the walk, that predict the performance of this quantum search algorithm provided certain spectral conditions are fulfilled. Our predictions are valid, for example, for (normalized) Hamiltonians whose spectral gap is considerably larger than $n^{-1/2}$. This allows us to derive necessary and sufficient conditions for optimal quantum search in this regime, as well as provide new examples of graphs where quantum search is sub-optimal. In addition, by extending this analysis, we are also able to show the optimality of quantum search for certain graphs with very small spectral gaps, such as graphs that can be efficiently partitioned into clusters. Our results imply that, to the best of our knowledge, all prior results analytically demonstrating the optimality of this algorithm for specific graphs can be recovered from our general results.
\end{abstract}
\date{\today}
\maketitle

\section{Introduction}

Quantum walks, the quantum analogue of classical random walks, find widespread applications in several areas of quantum information processing \cite{kempe2003quantum}. In particular, they are a universal model for quantum computation \cite{childs2009universal, childs2013universal} and are central to the design of several quantum algorithms \cite{ambainis2003quantum}.

The problem of finding a marked node in a graph, known as the spatial search algorithm, can be formulated as a continuous-time quantum walk (CTQW). In the original work by Childs and Goldstone \cite{childs2004spatial}, it was shown that search by CTQW can find a marked node in $\mathcal{O}(\sqrt{n})$ time \footnote{Throughout the article, we use a plethora of complexity theoretic notations which we briefly define now. $f(n)=\mathcal{O}(g(n))$, if there exists a positive constant $c$ such that $|f(n)|\leq c. g(n)$. If $f(n)=\mathcal{O}(g(n))$, then $g(n)=\Omega(f(n))$. Also $f(n)=o(g(n))$ if for all positive constants $k$, $|f(n)|<k.g(n)$. Consequently, if $f(n)=o(g(n))$, then $g(n)=\omega(f(n))$.} for certain graphs with $n$ nodes such as the complete graph, hybercube and $d$-dimensional lattices with $d>4$. This implied a quadratic speedup for the spatial search problem with respect to classical random walks for these graphs. However for lattices of $d\leq 4$, a full quadratic speedup is lost. Since then a plethora of results have been published exhibiting a $\mathcal{O}(\sqrt{n})$ running time of the Childs and Goldstone algorithm (henceforth referred to as the $\C\G$ algorithm) on certain specific graphs \cite{janmark2014global,novo2015systematic, chakraborty2016spatial, philipp2016continuous, wong2016quantum, chakraborty2017optimal, Boettcher_searchfractals_2017, wong2018quantum, glos2018optimal, rhodes2019quantum}. 

Although we state the general framework of the $\C\G$ algorithm in detail in Sec.~\ref{sec:prelim}, here we mention the algorithm briefly as this will aid the understanding of our contributions. Given a graph $G$ of $n$ nodes, the $\C\G$ algorithm involves evolving the following search Hamiltonian
\begin{equation}
\label{eqmain-search-ham-def}
H_{\mathrm{search}}=H_{\mathrm{oracle}}+ r H,
\end{equation}
where $r$ is a tunable parameter (the hopping rate of the quantum walk on $G$), $H$ is the Hamiltonian encoding the structure of $G$ (such as the graph's adjacency matrix or Laplacian) and $H_\mathrm{oracle}$ is the oracular Hamiltonian that singles out the marked node, which we shall denote as $\ket{w}$ \footnote{It is worth mentioning, that throughout the article we consider the scenario where only a single node is marked and analogous results for multiple marked nodes is an open problem in this framework.}. Typically, the algorithm commences from the highest eigenvector of $H$ which has a small overlap with $\ket{w}$ (say $\sqrt{\epsilon}$), and involves carefully choosing the value of $r$, such that evolving $H_{\mathrm{search}}$ for the minimum possible time, results in a state that has a large overlap with $\ket{w}$. It can be shown~\cite{farhi1998analog} that the algorithm is optimal if it can find the marked node with constant probability in $\Theta(\sqrt{\epsilon})$ time (in many cases $\epsilon=1/n$ as we discuss in Sec.~\ref{sec:prelim}). 

Most prior results on the optimality of the $\C\G$ algorithm, for particular graphs, have required an analysis specific to the underlying instance. For example in Ref.~\cite{janmark2014global}, the authors demonstrated, using degenerate perturbation theory that the $\C\G$ algorithm can find a marked node on a strongly regular graph in $\Oo(\sqrt{n})$ time, a graph that lacks global symmetry. Using similar techniques it was shown that a marked node can be found in optimal time on a graph with low algebraic connectivity \cite{meyer2015connectivity}. A long standing open problem has been to obtain the general conditions for the optimality of the algorithm and in particular to quantify the necessary and sufficient conditions a given graph must satisfy for the algorithm to be optimal.

A first attempt towards deriving sufficient conditions for the optimality of the algorithm was made in Ref.~\cite{chakraborty2016spatial} by connecting the graph spectral properties to the algorithmic running time. Namely, the authors demonstrated that the algorithm is optimal if the Hamiltonian encoding the graph structure,~i.e.\ $H$ has a constant spectral gap (without loss of generality, we assume $H$ to have eigenvalues in the interval $[0,1]$, see Sec.~\ref{sec:prelim} for details). However in the scenario where the spectral gap is no longer a constant, i.e, it decreases with the size of the graph, the results of the aforementioned work are not applicable. 

In our work, we significantly extend this result and provide the \emph{necessary and sufficient} conditions for the $\C\G$ algorithm to be optimal, for any $H$ obeying certain general assumptions (Sec.~\ref{sec:performance_non_degenerate}). The regime of validity of our results is defined by a \textit{spectral condition} which is obeyed, for example, when the spectral gap of $H$ (say $\Delta$) is sufficiently larger than the overlap of the initial state with the marked node, i.e.\ $\Delta\gg \sqrt{\epsilon}$. To the best of our knowledge, this condition is general enough to encompass most prior works predicting optimality of the $\C\G$ algorithm for specific graphs. Examples include the complete graph, hypercube \cite{childs2004spatial}, strongly regular graphs \cite{janmark2014global}, complete bipartite graphs \cite{novo2015systematic}, lattices of dimension greater than four~\cite{childs2004spatial}, Erd\"os-Renyi random graphs \cite{chakraborty2016spatial} or balanced trees \cite{philipp2016continuous} (for specific positions of the marked node).

To prove our optimality conditions, we first obtain general results regarding the best possible performance of the algorithm for any given graph obeying the previously mentioned validity regime. Precisely, we obtain general expressions, depending on the spectrum of $H$, for the optimal value of $r$, the maximum possible amplitude that can be obtained at the marked node and the time at which this amplitude is reached (Theorem \ref{lem_main:search-highest-estate}). The optimality conditions follow by imposing the maximum amplitude to be constant after a time $\sqrt{\epsilon}$. 

These predictions are, however, not valid for graphs with a sufficiently low spectral gap. Such low spectral gaps appear, for example, on graphs composed by highly connected clusters that are sparsely connected among each other, which find applications in spectral clustering \cite{von2007tutorial}. A simple example is the so-called \textit{joined complete-graph} of $n$ nodes: two complete graphs of $n/2$ nodes each, joined by a single edge between them (See Fig.~\ref{subfig:joined-complete}). If $H$ is defined by the normalized adjacency matrix of this graph, the spectral gap of $H$ is small enough to violate the spectral condition. However in Ref.~\cite{meyer2015connectivity}, using an analysis tailored to this particular instance, the authors showed that the $\C\G$ algorithm can find a marked node on this graph in $\Theta(\sqrt{n})$ time. 

Such instances are characterized by the following features in the spectrum of $H$: (i) A few of the highest eigenvalues are closely spaced (nearly degenerate), implying an extremely small spectral gap and (ii) A large gap between the closely spaced eigenvalues and the rest of the spectrum. We capture these properties precisely via new spectral conditions and provide, in Sec.~\ref{sec:search-degenerate-spectra}, a general theorem regarding the performance of quantum search on such graphs (Theorem~\ref{thm:performance_qd}). This leads to a sufficient condition that is able to predict optimal quantum search on the joined complete graph and other graphs with similar spectral properties.  


In Sec.~\ref{sec:quantum-walk-chessboard}, we provide an explicit example which compares and contrasts the applicability of Theorem \ref{lem_main:search-highest-estate} and Theorem \ref{thm:performance_qd}, respectively. Therein, we consider the quantum walk of a rook on a rectangular chessboard. This corresponds to the Cartesian product between two complete graphs, known as the \textit{Rook's graph} (See Fig.~\ref{figmain:chessboard}) \cite{moon1963line,hoffman1964line}. By altering the length and breadth of the chessboard (equivalently, by changing the size of the complete graphs) the spectral properties of the graph (namely, the spectral gap) can be changed. We identify different regimes of optimal and suboptimal quantum search, elucidating how the interplay between different choices of $r$ affect the algorithmic performance. Finally, we summarize and discuss upon the results of the article in Sec.~\ref{sec:discussion}.


\section{Preliminaries}
\label{sec:prelim}
First, we describe the framework of the $\C\G$ algorithm. Consider any graph $G$ with a set of $n$ vertices labelled $\{1,2,...n\}$ and a Hamiltonian $H$, which is an Hermitian matrix of dimension $n$ that encodes the connectivity of the underlying graph. In other words, we demand that $H$ is \textit{local}, i.e.\ its $(i,j)^{\mathrm{th}}$-entry is non-zero if and only if node $i$ (or $i^{\mathrm{th}}$ edge) is adjacent to node $j$ (or $j^{\mathrm{th}}$ edge) in $G$ (for example, $H$ could be proportional to the graph's adjacency matrix). Then, evolution under the Hamiltonian $H$ implements the continuous-time quantum walk on the graph that it encodes. Without loss of generality, it is assumed that $H$ has eigenvalues in the interval $[0,1]$ \footnote{This can be ensured by replacing $H$ with $(H/||H||+I)/2$, where $I$ is the identity matrix.}. 

As mentioned earlier the search Hamiltonian corresponding to the $\C\G$ algorithm is given by
$$
H_{\mathrm{search}}=H_{\mathrm{oracle}}+ r H.
$$
We require that $H_{\mathrm{oracle}}$ is local so that it perturbs the node $\ket{w}$ in a way that affects only vertices (or edges) in its vicinity. We will focus on the original formulation of the $\C \G$ algorithm where $H_{\mathrm{oracle}}=\ket{w}\bra{w}$ adds a local energy at node $\ket{w}$, leaving the remaining vertices unaffected. In fact, simulating this oracular Hamiltonian for a time $t$, corresponds to $\Oo(t)$-queries to the oracle of the Grover's search algorithm \cite{roland2003quantum}.
The steps of this algorithm are explained in Algorithm~\ref{algo-cg-general}. 

\RestyleAlgo{boxruled}
\begin{algorithm}[ht]
  \caption{$\C\G$ algorithm}\label{algo-cg-general}
Choose some $r> 0$ such that $H_{\mathrm{search}}=\ket{w}\bra{w}+rH$.
  \begin{itemize}
  \item[1)] Prepare the $1$-eigenstate of $H$.\\
  \item[2)] Evolve the state of $1)$ under $H_{\mathrm{search}}$ for some time $T$.
  \end{itemize}
\end{algorithm}
We note that our formulation of the $\CG$ algorithm is slightly more general than that of \cite{childs2004spatial}, where the authors consider the Hamiltonian $H_{\mathrm{search}}= - r L - \ket{w}\bra{w}$, such that $L$ is the Laplacian of the graph \cite{bollobas2013modern}, and choose the inital state $\ket{s}=n^{-1/2}\sum_i \ket{i}$ which is the 0-eigenstate of $L$. Our formulation becomes equivalent to that of \cite{childs2004spatial} if we set $H=I-\mathcal{L}$, where $\mathcal{L}$ is the normalized Laplacian and by suitably rescaling $r$.

The essential parameter in Algorithm~\ref{algo-cg-general} is the value of $r$, which has to be chosen judiciously so that the state $\ket{\Psi(T)}$ prepared after step 2) has a large overlap with the marked node
\begin{equation}
\alpha(T)=|\braket{w}{\Psi(T)}|,
\end{equation}   
for the minimum possible $T$. The marked node can then be obtained from this final state via a measurement on the state-space basis, or via amplitude amplification followed by measurement. 
\subsection{Running time of $\C\G$ algorithm}
To fully quantify the cost of running the $\CG$ algorithm, it is important to take into account not only the cost of evolving the search Hamiltonian for a given time, but also the cost to set-up the initial state and measure the final state. Furthermore, in some prior works ~\cite{childs2004spatial,childs2004spatial-dirac,childs2014spatial-crystal}  amplitude amplification has been used in conjunction with Hamiltonian evolution in order to find a marked node on lattices. To quantify the cost of such a procedure we need to introduce the cost of implementing the Grover oracle (the cost of evolving $H_{\text{oracle}}$ for constant time), which we denote as $\mathcal{C}_w$.  

We use the following notation for the different costs.
~\\
\begin{description}
\item[Setup cost $\mathcal{S}$] the cost of preparing the initial state of the algorithm ($1$-eigenstate of $H$). \footnote{The cost of reflecting on this state can be quantified as $2\mathcal{S}$. However, we shall omit constant factors for simplicity.}~\\
 \item[Time-evolution cost $\mathcal{T}$] The cost of implementing the time-evolution operator $e^{iTH_{\mathrm{search}}}$. A discrete simulation of this operator would require $O(T)$ queries to the Grover oracle \footnote{The results of \cite{berry2017exponential} on simulating continuous query models can be used to quantify the cost of implementing time-evolution of $H_{search}$ in terms of discrete queries to the Grover oracle as well as the cost of simulating $H$ and the error of the simulation.}.~\\
 \item[Measurement cost $\M$] The cost of performing a measurement in the state-space basis. 
\end{description}
 ~\\
Depending on the strategy used to obtain the marked node from the state $\ket{\Psi(T)}$, the overall cost can be quantified as follows (constant factors are omitted for simplicity):
\\~\\
 \textbf{i)~Amplitude Amplification}: Applying $1/\alpha(T)$ - rounds of the quantum amplitude amplification procedure \cite{brassard2002quantum} results in obtaining the marked node with constant probability. The overall running time of the $\C\G$ algorithm along with amplitude amplification is
\begin{equation}
\label{eqmain:search-time-with-amp-amp}
T_{\mathrm{search}}=\dfrac{1}{\alpha(T)}\left(\mathcal{S}+\mathcal{T}+\mathcal{C}_w\right)+\mathcal{M}.
\end{equation}
However, amplitude amplification is a discrete-time procedure, which implies that the overall algorithm is no longer continuous-time.

Furthermore, as evident from Eq.~\eqref{eqmain:search-time-with-amp-amp}, the setup cost plays a crucial role in the overall running time of the algorithm. In fact in prior works on the $\C\G$ algorithm, the \textit{Setup cost} $\mathcal{S}$ and the cost of making a measurement $\mathcal{M}$ have not been considered in order to compute the overall running time. 

It is important to guarantee that it is advantageous to run the quantum walk, as opposed to using only amplitude amplification on the initial state and bypassing the walk altogether. If the initial state of Algorithm \ref{algo-cg-general} has an overlap of $\sqrt{\epsilon}$ with $\ket{w}$,  the cost of the latter strategy is 
\begin{equation}
\label{eqmain:adv-amp-amp}
T_{AA}=\frac{1}{\sqrt{\epsilon}}\left (\mathcal{S} + \mathcal{C}_w\right)+\mathcal{M},
\end{equation}
where $\mathcal{C}_w$ is the cost of implementing the Grover oracle (evolving $H_{\mathrm{oracle}}$ for constant time). Clearly for the CTQW to be advantageous we need $T_{AA}$ to be larger than $T_{\mathrm{search}}$ from Eq.~\eqref{eqmain:search-time-with-amp-amp}.

In the case of the $\C\G$ algorithm, if the setup cost is reasonably large, such as for the applications considered in \cite{ambainis2005coins,ambainis2007quantum}, the aforementioned inequality is indeed satisfied and bypassing the quantum walk will invariably be disadvantageous. Hence, the choice of $r$ and $T$ should be such that the overhead due to amplitude amplification is as low as possible, or in other words $\alpha(T)$ is as large as possible. 
\\~\\

\textbf{ii)~Repetition}: By repeating the time-evolution followed by a measurement in the state-space basis $1/\alpha(T)^2$-times would result in obtaining the marked node with a high probability. The overall running time of the procedure in this case is
\begin{equation}
\label{eqmain:search-time-with-repetition}
T_{\mathrm{search}}=\dfrac{1}{\alpha(T)^2}\left(\mathcal{S}+\mathcal{T}+\mathcal{M}\right).
\end{equation}
Clearly, repeating the algorithm results in the overall running time being quadratically slower (with respect to $1/\alpha(T)$) as compared to that of amplitude amplification. However, if one assumes access only to the time-evolution of $H_{\mathrm{search}}$, then repeating this procedure is the only way to amplify the success probability. 
\\~\\

\subsection{Optimality of the algorithm}\label{sec:def_optimality} 
It is natural to ask, in this context, what is the minimum time needed to find the marked node for any Hamiltonian $H$. From the seminal work by Farhi and Gutmann \cite{farhi1998analog}, it is easy to obtain the following lower bound on the evolution time $T$ required to find $\ket{w}$
\begin{equation}
\label{eqmain:lower-bound-search}
T=\Omega\left(\dfrac{1}{\sqrt{\epsilon}}\right).
\end{equation} 
For vertex-transitive graphs, which informally means that there is no particular structure that allows to distinguish a node from any other node, we have that $\epsilon=1/n$, recovering the familiar Grover lower bound $T=\Omega(\sqrt{n})$ \footnote{For non vertex-transitive graphs, the structure can be such that certain particular nodes can be found faster than $\sqrt{n}$ time. For example, the central node on a star graph can be found in constant time. However, if any of the graph's nodes can be marked, the minimum time needed to find any node in a graph is lower bounded by $\sqrt{n}$ \cite{farhi1998analog}.}.

As such, throughout the article we shall say that the $\C\G$ algorithm is \emph{optimal} for a given graph $G$ if Algorithm \ref{algo-cg-general} results in a state that has a constant overlap with $\ket{w}$ after evolving for a time that matches the aforementioned lower bound, i.e.\ $T=\Theta\left(1/\sqrt{\epsilon}\right)$. In such a case, the overall search time would  scale as
\begin{equation}
T_{\mathrm{search}}=\mathcal{S}+\Theta\left(\dfrac{\mathcal{C}_w}{\sqrt{\epsilon}}\right)+\M,
\end{equation}
assuming the cost of implementing the walk evolution for time $T=\Theta\left(1/\sqrt{\epsilon}\right)$ is dominated by the cost $\mathcal{C}_w$ of implementing the oracle, i.e.,
\begin{equation}
    \mathcal{T}=\Theta\left(\dfrac{\mathcal{C}_w}{\sqrt{\epsilon}}\right).
    \end{equation}
In this scenario, running the walk is advantageous as compared to simply doing amplitude amplification on the initial state (Eq.~\eqref{eqmain:adv-amp-amp}) when the set-up cost is considerably larger than the cost of implementing the oracle. 

It is worth noting that in order to quantify a speedup, one can also compare the running time of quantum spatial search with the time required by classical random walks to solve the same problem. In fact, the time required by a classical random walk to find a marked node on any graph, known as the \textit{hitting time}, is bounded as follows:
\begin{equation}
\label{eqmain:hitting-time-bounds}
\dfrac{1}{\epsilon}\leq \mathrm{HT}(w) \leq \dfrac{1}{\Delta\epsilon},
\end{equation} 
where $\Delta$ is the spectral gap of the operator defining the random walk (such as the normalized adjacency matrix or the graph Laplacian). 

In fact, it has been established that discrete-time quantum walk search algorithms \cite{magniez2011search,krovi2016quantum,ambainis2019quadratic} as well as the recent continuous-time quantum walk search algorithm we proposed \cite{chakraborty2018finding}, can find a marked node on any graph in square root of the hitting time, resulting in a generic quadratic advantage. However, such algorithms require a larger Hilbert space and can be seen as quantum walks on the edges of the underlying graph. As such, in this article we shall also compare the running time of the $\C\G$ algorithm with the \textit{hitting time} of classical random walks, towards identifying the regimes for which a quadratic speed-up can be obtained as well as the limitations of this framework for quantum search.

\section{Performance of the $\mathbb{\C\G}$ algorithm}
\label{sec:performance_non_degenerate}
In this section we derive the main results characterizing the performance of the $\CG$ algorithm.  Let $H$ have eigenvalues $0\leq \lambda_1\leq \lambda_2\leq\cdots\lambda_{n-1}<\lambda_n=1$ with the corresponding eigenvectors, $\ket{v_1},\cdots,\ket{v_n}$ such that 
\begin{equation}
H\ket{v_i}=\lambda_i\ket{v_i}.
\end{equation}
Also let the gap between the two highest eigenvalues of $H$ (spectral gap) be given by  
\begin{equation}
\Delta=1-\lambda_{n-1}.
\end{equation}
It will be convenient to express the marked node in the basis of the eigenstates of $H$ as 
\begin{equation}
\label{eqmain:sol-in-eigenbasis}
\ket{w}=\sum_{i=1}^{n}a_i\ket{v_i}
\end{equation}
and define the following set of parameters 
\begin{equation}
\label{eqmain:Sk}
S_k=\sum_{i=1}^{n-1}\dfrac{|a_i|^2}{(1-\lambda_i)^k},
\end{equation}
for integer $k\geq 1$. These parameters depend only on the spectral properties of the graph and the position of the marked node and turn out to be crucial to understanding the algorithmic performance, as it is clear, for example,  in the studies of quantum search on lattices~\cite{childs2004spatial} and fractals \cite{Boettcher_searchfractals_2017}. We note also that for vertex-transitive graphs these parameters depend only on the eigenvalues of $H$, as all the probabilities $|a_i|^2=1/n$.

Furthermore, we impose the following \emph{spectral condition} that defines the regime of validity of our analysis
\begin{equation}\label{eq:spectral_con}
\sqrt{\epsilon} < c \min\left\{\frac{S_1 S_2}{S_3},\Delta\sqrt{S_2}\right\},
\end{equation}
where $c$ is a small positive constant. The reason why we need to impose this condition will become clear in Sec.~\ref{sec:criticalpoint}. In a nutshell, this condition ensures that we can bound the error in our perturbative analysis and furthermore, that the additive error we obtain in the final amplitude at the marked node is small enough for our predictions to be meaningful.

In subsection~\ref{subsec:applicability-of-lemma-1} we discuss the generality of this condition and prove that it is fulfilled for any graph where $\sqrt{\epsilon} \leq c \Delta$. However, it is more general than that since it also includes the critical case of the 4d-lattice, where both $\sqrt{\epsilon}$ and $\Delta$ scale as $\Theta(1/\sqrt{n})$.  

Our main results regarding the performance of the algorithm are the following. In subsection~\ref{sec:criticalpoint}, we demonstrate that the optimal choice for $r$ is $r=S_1$,  provided the spectral condition from Eq.~\eqref{eq:spectral_con} is respected. In this case we show that the maximum amplitude at the solution is reached at time
\begin{equation}
\label{eqmain:evolution-time-cg-algorithm}
T=\Theta\left(\dfrac{1}{\sqrt{\epsilon}}\dfrac{\sqrt{S_2}}{S_1}\right)
\end{equation} 
and is given by 
\begin{equation}\label{eq:maxamp}
\nu \approx \dfrac{S_1}{\sqrt{S_2}}.
\end{equation}
Furthermore, we show that essentially the same behavior is maintained if we choose $r$ within a window of $|r-S_1|= O(\sqrt{\epsilon S_2})$. If $r$ is chosen outside this interval, we show in Sec.~\ref{sec:failure_critpoint} that the maximum amplitude reached is  $o(S_1/\sqrt{S_2})$, independently of the evolution time we choose. This implies that $\Theta(S_1/\sqrt{S_2})$ is the maximum amplitude achievable for any time and any choice of $r$. 

Consequently, we can draw the following \emph{necessary and sufficient condition} for optimal quantum search (in the sense discussed in Sec.~\ref{sec:def_optimality}), within the regime of validity of our analysis.
\begin{theorem}[Optimality of quantum search]
\label{thm:optimalityCG}
Let $H$ be such that the spectral condition from Eq.~\eqref{eq:spectral_con} is fulfilled. Then the $\CG$ algorithm is optimal iff $S_1/\sqrt{S_2}=\Theta(1)$.
\end{theorem}
The proof of this result follows directly from the statements above. If $S_1/\sqrt{S_2}=\Theta(1)$, choosing $r$ sufficiently close to $S_1$ ensures that we obtain a constant amplitude at the marked node after $T= \Theta\left(1/\sqrt{\epsilon}\right)$ (see Eqs. \eqref{eqmain:evolution-time-cg-algorithm} and \eqref{eq:maxamp}), matching the lower bound in Eq.~\eqref{eqmain:lower-bound-search}. On the other hand, since $\Theta(S_1/\sqrt{S_2})$ is the maximum amplitude achievable, if we have that $S_1/\sqrt{S_2}=o(1)$ the algorithm is never optimal. 

With this necessary and sufficient condition, many hitherto published results showing that this algorithm is optimal for specific graphs can be recovered , without the need to do a graph specific analysis \cite{childs2004spatial, janmark2014global,novo2015systematic,wong2016quantum,wong2016quantum2,philipp2016continuous,
wong2016laplacian,wong2016engineering,chakraborty2016spatial, chakraborty2017optimal,glos2018vertices,glos2018optimal,
wong2018quantum}. For example, it is possible to see, from the fact that $S_1/\sqrt{S_2}\geq \sqrt{\Delta}$ (see Lemma  \ref{lem:nu-bound-1}), that search is optimal for any Hamiltonian $H$ with a constant spectral gap, and thus recover the main result from Ref.~\cite{chakraborty2016spatial}. This encompasses graphs such as Erd\"os-Renyi random graphs, complete bipartite graphs or strongly regular graphs. Additionally, our results also predict optimality for graphs such as hypercubes, lattices of dimension greater than four even though they do not exhibit a constant spectral gap. 

One can wonder whether a simpler and more intuitive sufficient condition for optimality can be derived from Theorem~\ref{thm:optimalityCG}. To our knowledge, all previously known examples of graphs whose spectral gap is large enough compared to $\sqrt{\epsilon}$ can be searched by quantum walk in optimal time (e.g. lattices of dimension $d\geq5$), so one could think that $\sqrt{\epsilon}\ll \Delta$ is a sufficient condition for optimal quantum search. We show explicitly that this is not the case -- there exist graphs for which the spectral condition is satisfied and the spectral gap is such that $\sqrt{\epsilon}\ll \Delta$ but nevertheless the value of  $S_1/\sqrt{S_2}$ decreases with the size of the graph which implies suboptimality. This is the case, for example, for the normalized adjacency matrix of the Rook's graph in some regime (see Sec.~\ref{sec:quantum-walk-chessboard}).  In fact, this example shows that this quantum walk algorithm can also be slower than the square root of the hitting time of the corresponding classical walk.

\subsection{Performance of quantum search at the critical point ($r\approx S_1$)}\label{sec:criticalpoint}
Here we present one of our main results, which characterizes the performance of quantum search when the parameter $r$ is close to its optimal value. 
~\\
\begin{theorem}
\label{lem_main:search-highest-estate}
Let $H$ be such that the spectral condition of Eq.~\eqref{eq:spectral_con} is obeyed, with $S_k$ defined as in Eq.~\eqref{eqmain:Sk}. By choosing $r=S_1$ and 
$$T=\Theta\left(\dfrac{1}{\sqrt{\epsilon}}\dfrac{\sqrt{S_2}}{S_1}\right),$$
Algorithm \ref{algo-cg-general} prepares a state $\ket{f}$ such that $\nu=|\braket{w}{f}|=\Theta\left(S_1/\sqrt{S_2}\right)$.
\end{theorem}
~\\
\begin{proof} 
As defined previously, the Hamiltonian $H$ has eigenvalues $\lambda_i$ and corresponding eigenvectors $\ket{v_i}$. We denote each eigenvalue of $rH$ as $\lambda'_{i}:=r\lambda_{i}$.
First, we express the solution state $\ket{w}$ in terms of the eigenstates of $H$. We have
\begin{equation}
\label{eqmain:solution_expanded}
\left|w\right\rangle =\sum_{i=1}^{n}a_{i}\left|v_{i}\right\rangle, 
\end{equation}
such that $|a_n|=\sqrt{\epsilon}$. Now we find the condition for which a quantum state $\ket{\psi}$ 
defined as 
\begin{equation}
\left|\psi\right\rangle =\sum_{i=1}^{n} b_{i}\left|v_{i}\right\rangle,
\end{equation}
is an eigenstate of $H_{\mathrm{search}}=rH+H_{\mathrm{oracle}}$. That is,
\begin{align}
H_{\mathrm{search}}\left|\psi\right\rangle  & =E\ket{\psi}\\
\implies\sum_{i}\lambda_{i}'b_{i}\left|v_{i}\right\rangle +\underbrace{\left\langle w|\psi\right\rangle }_{=:\gamma}\left|w\right\rangle &= E\ket{\psi}\\
\implies\sum_{i}\left(\lambda_{i}'b_{i}+\gamma a_{i}\right)\left|v_{i}\right\rangle &=\sum_{i}E b_{i}\left|v_{i}\right\rangle.
\end{align}
This implies that 
\begin{equation}
\label{eq:coeff-estates-Hsearch}
b_{i}=\frac{\gamma a_{i}}{E-\lambda'_{i}}.
\end{equation}
Note that 
$$\gamma=\left\langle w|\psi\right\rangle =\sum_{i}a^*_{i}b_{i}$$
where we substitute for $b_{i}$ to get
\begin{equation}
\label{eq:condition_new_eigenvalue}
1=\sum_{i}\dfrac{|a_{i}|^{2}}{E-\lambda_{i}'}.
\end{equation}
This equation gives us the condition for $E$ to be an eigenvalue of $H_{\text{search}}$. It can be seen that the RHS of \eqref{eq:condition_new_eigenvalue} is a monotonically decreasing function of $E$ within each interval $]\lambda'_{i-1},\lambda'_{i}[$ and $\lambda'_{i}$ are poles of this function. This guarantees that each of these intervals, as well as as the interval $]\lambda'_n, +\infty[$, contains exactly one eigenvalue. 

We are interested in finding the two largest eigenvalues of $H_{\text{search}}$. We choose $r=S_1$ and will look for solutions of Eq.~\eqref{eq:condition_new_eigenvalue} of the form $E=\lambda'_n+\delta$, within the interval $|\delta| < c' S_1 \Delta $ for some small constant $c'$. Indeed, we will demonstrate that there are two solutions within this interval.

To show this, we rewrite Eq.~\eqref{eq:condition_new_eigenvalue} in terms of $\delta$ and choose $r=S_1$ to obtain
\begin{equation}
\label{eq:condition-new-evalue-2}
\frac{\epsilon}{\delta}+\sum_{i<n}\frac{|a_i|^2}{S_1\Delta_i+\delta}=1,
\end{equation}
where $\Delta_i=\lambda_n-\lambda_i$. Finding solutions of this equation is equivalent to finding the zeros of a function $F(\delta)$ which can be written as
\begin{align}
F(\delta)&=\dfrac{\epsilon}{\delta}+\sum_{i<n}\dfrac{|a_i|^2}{S_1\Delta_i+\delta}-1\\
		 &=\dfrac{\epsilon}{\delta}+\sum_{i<n}\dfrac{|a_i|^2}{S_1\Delta_i}\left(1+\sum_{k=1}^{\infty}\dfrac{(-\delta)^k}{S^k_1\Delta^k_i}\right)-1\\
		 &= \dfrac{\epsilon}{\delta}-\dfrac{S_2\delta}{S_1^2}+\sum_{i<n}\dfrac{|a_i|^2\delta^2}{S_1^3\Delta_i^3}\sum_{k=0}^{\infty}
		 \dfrac{(-\delta)^k}{\Delta^k_i S_1^k} \\
		  &=\dfrac{\epsilon}{\delta}\left\{1-\dfrac{S_2\delta^2}{S_1^2\epsilon}+f(\delta)\right\}\label{eq:Fdelta},
\end{align}
The term $f(\delta)$ can be seen as an error term that can be bounded as 
\begin{align}
\label{eq:error-bound-taylor}
|f(\delta)|\leq\dfrac{S_3|\delta|^3}{S^3_1\epsilon}\dfrac{1}{1-\frac{|\delta|}{S_1\Delta}}\leq \dfrac{S_3|\delta|^3}{S^3_1\epsilon}\dfrac{1}{1-c'}.
\end{align}
If this error term was neglected, the function $F(\delta)$ would have zeros at $\pm \delta_0$, where
\begin{equation}\label{eq:def_delta0}
\delta_0=\sqrt{\epsilon} \dfrac{S_{1}}{\sqrt{S_2}}.
\end{equation}
We will see that the presence of the term $f(\delta)$ introduces a relative error in these solutions i.e., there are two solutions $\delta_{\pm}$ in the intervals 
\begin{align}
\delta_+&\in\left[(1-\eta)\delta_0,(1+\eta)\delta_0\right] \label{eq:interval_deltaplus}\\
\delta_-&\in\left[-(1+\eta)\delta_0,-(1-\eta)\delta_0\right]\label{eq:interval_deltaminus},
\end{align} 
where the relative error is given by 
\begin{equation}
\label{eq:eta-ratio}
\eta=\dfrac{S_3\sqrt{\epsilon}}{S^{3/2}_2}.
\end{equation}
To demonstrate this let us focus on the interval given by Eq.~\eqref{eq:interval_deltaplus} and show that $F(\delta)$ has a zero in this interval (an analogous derivation can be done for the other interval in Eq.~\eqref{eq:interval_deltaminus}). If we take $\delta_+=\delta_0(1+\eta')$, where $|\eta'|\leq \eta$ we can bound $|f(\delta_+)|$ as  
\begin{align}
\label{eq:error-bound2}
|f(\delta_+)|\leq\dfrac{S_3\delta_0^3(1+\eta)^3}{S^3_1\epsilon(1-c')}\leq \dfrac{\eta(1+\eta)^3}{(1-c')},
\end{align}    
where we used the definitions from Eqs.~\eqref{eq:def_delta0} and \eqref{eq:eta-ratio}. Note that the spectral condition imposed in Eq.~\eqref{eq:spectral_con} guarantees that $\eta$ is small. Using this condition we can show that 
\begin{equation}\label{eq:bound_eta}
    \eta\leq c \frac{S_1}{\sqrt{S_2}}\leq c,
\end{equation}
where we also used the fact that  $\frac{S_1}{\sqrt{S_2}}\leq 1$ which is demonstrated later in Lemma~\ref{lem:nu-bound-1}.

On the other hand, from \eqref{eq:Fdelta} we have that
\begin{align}\label{eq:Fdeltaplus}
 F(\delta_+)=\dfrac{\epsilon}{\delta_+}\left\{2\eta'+\eta'^2+f(\delta_+)\right\}.   
\end{align}
Given the bound \eqref{eq:error-bound2}, we see that the RHS of  \eqref{eq:Fdeltaplus} is positive if $\eta'=\eta$ and negative if $\eta'=-\eta$, provided $c$ and $c'$ are sufficiently small. This shows that indeed $\delta_+$ is in the interval from Eq.~\eqref{eq:interval_deltaplus}. The same reasoning can be used to show that $\delta_-$ belongs to the interval in Eq.~\eqref{eq:interval_deltaminus}.

We can now verify the validity of the assumption $\delta_+\leq c' S_1 \Delta$, for some small constant $c'$, which was necessary to obtain the bound in Eq.~\eqref{eq:error-bound-taylor}. We have that 
\begin{align}
    \delta_+&\leq \delta_0 (1+|\eta|)\\
    &\leq \frac{S_1}{\sqrt{S_2}}\sqrt{\epsilon}(1+c)\\
    &\leq c (1+c) S_1 \Delta,
\end{align}
where in the second step we used Eqs.~\eqref{eq:bound_eta} and \eqref{eq:def_delta0} and in the last step we used the spectral condition \eqref{eq:spectral_con}. Hence, for sufficiently small $c$ the condition $\delta_+\leq c' S_1 \Delta$ is verified. 

Now that we have obtained two approximate solutions to Eq.~\eqref{eq:condition_new_eigenvalue} $E_{\pm}=\lambda'_n+\delta_{\pm}$, we proceed to compute the overlap of the corresponding eigenstates $\ket{\psi_{\pm}}$ with the marked node. From the normalization condition of the eigenstates of $H_{\mathrm{search}}$, we have $\sum |b_{i}|^{2}=1$. So from Eq.~\eqref{eq:coeff-estates-Hsearch}, we can obtain the following equation for $\gamma_{\pm}=\braket{w}{\psi_{\pm}}$
\begin{align}
&\sum_{i}\frac{|\gamma_\pm a_{i}|^2}{\left(E_\pm-\lambda'_{i}\right)^{2}}=1\\
\implies &|\gamma_\pm|^2=\left[\frac{\epsilon}{\delta_\pm^{2}}+\sum_{i<n}\frac{|a_{i}|^{2}}{\left(E_\pm-\lambda'_{i}\right)^{2}}\right]^{-1}\\
\implies |\gamma_\pm|^2 &=\left[\frac{\epsilon}{\delta_\pm^{2}}+\sum_{i<n}\frac{|a_{i}|^{2}}{S_1^2\Delta_i^2\left(1+\frac{\delta_\pm}{S_1\Delta_i}\right)^{2}}\right]^{-1}.
\end{align} 

Replacing the values of $\delta^\pm$ and neglecting terms scaling as $\Theta\left(\eta^2\right)$ we obtain
\begin{align}
&|\gamma_\pm|^2=\dfrac{S_1^2}{2S_2}\left(1+\Theta\left(\eta\right)\right)\\
\implies & |\gamma_\pm|=\dfrac{S_1}{\sqrt{2S_2}}\left(1+\Theta\left(\eta\right)\right)      
\end{align}
Without loss of generality we can choose the eigenbasis of $H_{search}$ such that the overlaps $\gamma_{\pm}$ are positive. 
Furthermore, we can calculate the values of $b^\pm_{n}=\braket{\psi_{\pm}}{v_n}$ from Eq.~\eqref{eq:coeff-estates-Hsearch}, which yields
\begin{equation}
b^\pm_{n}=\gamma_\pm\frac{a_{n}}{\delta_\pm}.
\end{equation}
Substituting the values of $\delta_\pm$ and $\gamma_\pm$, we obtain that, \begin{align}
b^\pm_n&=\pm \dfrac{1}{\sqrt{2}}\left(1+\Theta(\eta)\right).
\end{align}
So the starting state can be written as
\begin{equation}
\left|v_{n}\right\rangle = \dfrac{\ket{\psi_+}-\ket{\psi_-}}{\sqrt{2}}+\ket{\Phi},
\end{equation}
where $\ket{\Phi}$ is an unnormalized quantum state such that $\nrm{\ket{\Phi}}\leq \Theta\left(\eta\right)$.

So evolving $\ket{v_n}$ under $H_{\mathrm{search}}$ for a time $t$ results in  
\begin{align}
&e^{-iH_{\mathrm{search}}t}\left|v_{n}\right\rangle\\
&=\frac{1}{\sqrt{2}}e^{-i\lambda'_{n}t}\left(e^{-i\delta_+ t}\left|v_{+}\right\rangle -e^{-i\delta_- t}\left|v_{-}\right\rangle \right)+\Theta\left(\eta\right).
\end{align}
Thus after a time $T=\frac{\pi}{2\delta_0}=\Theta(\frac{1}{\sqrt{\epsilon}}\frac{\sqrt{S_2}}{S_1})$, up to a global phase, we end up in the state
$$\ket{f}=\dfrac{\ket{v_+}+\ket{v_-}}{\sqrt{2}}+\ket{\Phi'},$$
such that $\nrm{\ket{\Phi'}}\leq \Theta\left(\eta\right)$. The overlap of $\ket{f}$ with the solution state is given by 
\begin{align}
\nu&=|\braket{w}{f}|\\
	   &=\dfrac{S_1}{\sqrt{S_2}}\left(1+\Theta\left(\eta\frac{\sqrt{S_2}}{S_1}\right)\right)=\Theta\left(\dfrac{S_1}{\sqrt{S_2}}\right),
\end{align}
where we have used the spectral condition \eqref{eq:spectral_con} which ensures  that
$$
\eta\dfrac{\sqrt{S_2}}{S_1}=\dfrac{\sqrt{\epsilon}S_3}{S_1 S_2}\leq c.
$$
\end{proof}
~\\
Subsequently, we show that essentially the same behavior is maintained if we choose any $r$ within a small enough interval around $S_1$.
~\\
\begin{theorem}
\label{lem_main:robustness_r}
Let $H$ be such that the spectral condition of Eq.~\eqref{eq:spectral_con} is obeyed and  $r$ be chosen such that
\begin{equation}\label{eq:ropt_deviation}
|r-S_1|\leq |\beta| \sqrt{\epsilon S_2},    
\end{equation} 
for some small constant $\beta$ such that $|\beta|\ll 1$.
After a time
$$T=\Theta\left(\dfrac{1}{\sqrt{\epsilon}}\dfrac{\sqrt{S_2}}{S_1}\right),$$
Algorithm \ref{algo-cg-general} prepares a state $\ket{f}$ such that $\nu=|\braket{w}{f}|=\Theta\left(S_1/\sqrt{S_2}\right)$.
\end{theorem}
~\\
\begin{proof}
The proof of this result follows from the fact that, for any  $r$ within this interval, we still have that the value of $|\delta_{\pm}|=\Theta(\delta_0)$. More precisely, by rewriting  Eq.~\eqref{eq:Fdelta} for arbitrary $r$ we have
\begin{equation}
    F(\delta)= \frac{\epsilon}{\delta}\left\{1+\frac{\delta }{\epsilon }\left(\frac{S_1}{r}-1\right)- \frac{S_2\delta^2}{r^2\epsilon}\right\}
\end{equation}
which will have two zeros in the intervals 
\begin{align}
\delta_+&\in\left[(1-\eta)\delta^{(+)}_0,(1+\eta)\delta_0^{(+)}\right] \label{eq:interval_deltaplus}\\
\delta_-&\in\left[-(1+\eta)\delta^{(-)}_0,-(1-\eta)\delta^{(-)}_0\right]\label{eq:interval_deltaminus},
\end{align}
where 
\begin{equation}
    \delta_0^{(\pm)}=\left|\frac{S_1\sqrt{\epsilon}}{\sqrt{S_2}}\left(\frac{\beta}{2}\pm \sqrt{1+\frac{\beta^2}{4}}\right)\right|+\mathcal{O}(\epsilon)
\end{equation}
which is of the same order of the value $\delta_0$ from Eq.~\eqref{eq:def_delta0}. Hence, with analogous arguments to those used in the proof of Theorem~\ref{lem_main:search-highest-estate} we conclude that a deviation to the optimal value of $r$ as in Eq.~\eqref{eq:ropt_deviation}, only changes the maximum amplitude and the evolution time needed to reach this amplitude by constant factors. 
\end{proof}
~\\
\subsection{Failure of the algorithm away from $\mathbf{r\approx S_1}$}\label{sec:failure_critpoint}
Previously, we have established that Algorithm \ref{algo-cg-general} prepares a state with an overlap of $\Theta(S_1/\sqrt{S_2})$ with the marked vertex for any choice of $r$ within the interval
\begin{equation}
\label{eq:robust-range-of-r}
r^* \in \left[S_1-\Theta\left(\sqrt{S_2\epsilon}\right), S_1+\Theta\left(\sqrt{S_2\epsilon}\right) \right],
\end{equation}
after a time 
$$T=\Theta\left(\dfrac{1}{\sqrt{\epsilon}}\dfrac{\sqrt{S_2}}{S_1}\right).$$

In this section we prove that for any choice of $r$ outside the window mentioned in Eq.~\eqref{eq:robust-range-of-r}, the amplitude of the algorithm is less than $S_1/\sqrt{S_2}$, irrespective of $T$. 
~\\
\begin{theorem}
\label{lem:sub-optimality-for-any-r}
For any $r\geq 0$, such that $r \notin r^*$, 
$$
|\braket{w}{e^{-iH_{\mathrm{search}}T}|v_n}|\leq o\left(\dfrac{S_1}{\sqrt{S_2}}\right),
$$
$\forall T\geq 0$.
\end{theorem}
~\\
\begin{proof}
In order to derive this, we require all the eigenvalues and eigenvectors of $H_{\mathrm{search}}$. As such we consider that $H_{\mathrm{search}}$ has eigenvalues $E_n > E_{n-1}>\cdots \geq E_1$, such that $H_{\mathrm{search}}\ket{\psi_i}=E_i\ket{\psi_i}$. As before, we consider that the eigenvectors and corresponding eigenvalues of $H$ are $\lambda_i$ and $\ket{v_i}$, respectively.

Then for $1\leq \alpha \leq n$ let
\begin{equation}
\ket{\psi_\alpha}=\sum_{i=1}^n b^{(\alpha)}_i\ket{v_i},
\end{equation}
and  define
\begin{equation}
\label{eq:gamma-alpha}
\gamma_\alpha =\braket{w}{\psi_\alpha}.
\end{equation}

So, by using the fact that $H_{\mathrm{search}}\ket{\psi_i}=E_i\ket{\psi_i}$, we obtain 
\begin{equation}
\label{eq:b-i-alpha}
b^{(\alpha)}_i=\dfrac{\gamma_\alpha a_i}{E_\alpha - r \lambda_i},~~1\leq\alpha\leq n.
\end{equation} 
For all $1\leq \alpha \leq n$, we use the definition of $\gamma_\alpha$ in Eq.~\eqref{eq:gamma-alpha} and the expression for $b^{(\alpha)}_i$ in Eq.~\eqref{eq:b-i-alpha} to obtain 
\begin{align}
\label{eq:eigenvalue-condition-alpha}
F(E_{\alpha}):=\sum_{i=1}^n \dfrac{|a_i|^2}{E_\alpha-r\lambda_i}=1.
\end{align}

From the normalization condition, $\sum_i |b^{(\alpha)}_i|^2=1$, for every $\alpha$, we also obtain that
\begin{equation}\label{eq:gamma-alpha2}
\dfrac{1}{|\gamma_\alpha|^2}=\sum_{i=1}^n \dfrac{|a_i|^2}{(E_\alpha-r\lambda_i)^2}.
\end{equation}

Now,
\begin{align}
\braket{w}{e^{-iH_{\mathrm{search}}t}|v_n}&=\sum_{\alpha}e^{-iE_\alpha t}\braket{w}{\psi_\alpha}\braket{\psi_\alpha}{v_n}\\
										  &=\sqrt{\epsilon}\sum_{\alpha}e^{-iE_\alpha t}\gamma_\alpha b^{(\alpha) *}_n\\                                        
\label{eq:final-amplitude-any-t}                                          
                                          &=\sqrt{\epsilon}\sum_{\alpha} \dfrac{|\gamma_\alpha|^2 e^{-iE_\alpha t}}{E_\alpha -r},
\end{align}
where in the last line we have replaced the value of $b^{(\alpha)}_n$ from Eq.~\eqref{eq:b-i-alpha}.

Note that from the condition that the amplitude at $t=0$ is $\sqrt{\epsilon}$ we obtain 
\begin{equation}
\label{eq:initial-amplitude}
\sum_{\alpha=1}^n \dfrac{|\gamma_\alpha|^2}{E_{\alpha}-r}=1.
\end{equation}

Now we shall consider the following two cases, each of which we treat differently: (i) When $r>r^*$ and (ii) $r<r^*$. 
\\~\\
\textbf{(i) $\mathbf{r>r^*}$}: We first use Eq.~\eqref{eq:final-amplitude-any-t} and Eq.~\eqref{eq:initial-amplitude} to obtain that
\begin{equation}
\label{eq:upper-bound-amplitude}
|\braket{w}{e^{-iH_{\mathrm{search}}t}|v_n}|\leq \dfrac{2\sqrt{\epsilon}|\gamma_n|^2}{E_n-r}-\sqrt{\epsilon}.
\end{equation}

Let $E_n=r+\delta_+$ and $\Delta_j=1-\lambda_j$. From Eq.~\eqref{eq:gamma-alpha2} we have
\begin{align}
 \dfrac{1}{|\gamma_n|^2}&\geq \dfrac{\epsilon}{\delta^2_+}
\label{eq:upper-bound-gamma-n}\\
\implies |\gamma_n|^2&\leq \dfrac{\delta^2_+}{\epsilon}.
\end{align}

Substituting this into Eq.~\eqref{eq:upper-bound-amplitude}, we get
\begin{equation}
\label{eq:upper-bound-amplitude-2}
|\braket{w}{e^{-iH_{\mathrm{search}}t}|v_n}|\leq \dfrac{2\delta_+}{\sqrt{\epsilon}}
\end{equation}

Next using the fact that $F(E_n)=1$ (Eq.~\eqref{eq:eigenvalue-condition-alpha}), we obtain an upper bound on $\delta_+$ as follows
\begin{align}
&\dfrac{\epsilon}{\delta_+}+\sum_{j=1}^{n-1}\dfrac{|a_j|^2}{E_n-r+r\Delta_j}=1\\
&\implies \dfrac{\epsilon}{\delta_+}+\dfrac{S_1}{r}>1\\
\label{eq:upper-bound-delta-+}
&\implies \delta_+ < \dfrac{\epsilon r}{r-S_1}.
\end{align}
Next we substitute the upper bound on $\delta_+$ into Eq.~\eqref{eq:upper-bound-amplitude-2} to obtain 
\begin{equation}
|\braket{w}{e^{-iH_{\mathrm{search}}t}|v_n}|\leq \dfrac{2\sqrt{\epsilon}r}{S_1-r}.
\end{equation} 
For any $r=S_1+\omega\left(S_2\sqrt{\epsilon}\right)$, we indeed obtain that
\begin{equation}
|\braket{w}{e^{-iH_{\mathrm{search}}t}|v_n}|\leq o\left(\dfrac{S_1}{\sqrt{S_2}}\right).
\end{equation}
\\~\\
(ii) \textbf{$\mathbf{r<r^*}$:~}In this region, the proof is similar in spirit to the case where $r>r^*$. Here we can bound the amplitude by bounding the value of $\delta_-$ where $E_{n-1}=r-\delta_-$.

In fact as before, using Eq.~\eqref{eq:final-amplitude-any-t} and Eq.~\eqref{eq:initial-amplitude} to obtain that
\begin{equation}
\label{eq:upper-bound-amplitude-left}
|\braket{w}{e^{-iH_{\mathrm{search}}t}|v_n}|\leq \dfrac{2\sqrt{\epsilon}|\gamma_{n-1}|^2}{\delta_-}+\sqrt{\epsilon}.
\end{equation}
From Eq.~\eqref{eq:gamma-alpha}, it is easy to obtain that
\begin{align}
\label{eq:upper-bound-gamma-n-left}
|\gamma_{n-1}|^2\leq \dfrac{\delta^2_-}{\epsilon}.
\end{align}

This gives us
\begin{equation}
\label{eq:upper-bound-amplitude-2-left}
|\braket{w}{e^{-iH_{\mathrm{search}}t}|v_n}|\leq \dfrac{2\delta_-}{\sqrt{\epsilon}}+\sqrt{\epsilon}
\end{equation}

Here we use the fact that $F(E_{n-1})=1$ to obtain an upper bound on $\delta_-$. We have,
\begin{align}
&-\dfrac{\epsilon}{\delta_-}+\sum_{j=1}^{n-1}\dfrac{|a_j|^2}{-\delta_-+r\Delta_j}=1\\
&\implies -\dfrac{\epsilon}{\delta_-}+\dfrac{S_1}{r}<1~~~~~~[\because~\delta_-< r\Delta]\\
\label{eq:upper-bound-delta--}
&\implies \delta_- < \dfrac{\epsilon r}{S_1-r}.
\end{align}
We now substitute the upper bound on $\delta_-$ into Eq.~\eqref{eq:upper-bound-amplitude-2-left} to obtain 
\begin{equation}
|\braket{w}{e^{-iH_{\mathrm{search}}t}|v_n}|\leq \dfrac{2\sqrt{\epsilon}r}{r-S_1}+\sqrt{\epsilon}.
\end{equation} 
Thus any $r=S_1-\omega\left(S_2\sqrt{\epsilon}\right)$, we indeed obtain that
\begin{equation}
|\braket{w}{e^{-iH_{\mathrm{search}}t}|v_n}|\leq o\left(\dfrac{S_1}{\sqrt{S_2}}\right),
\end{equation}
where in the last line we use the fact that we are in a regime where $\sqrt{\epsilon}=o\left(S_1/\sqrt{S_2}\right)$.
\\~\\
This concludes the proof.
\end{proof}


\subsection{Validity of the spectral condition and implications to algorithmic performance}
\label{subsec:applicability-of-lemma-1}
We have seen that the maximum amplitude at the marked node is determined by the ratio $S_1/\sqrt{S_2}$ and hence it is important to obtain upper and lower bounds on this quantity, given any $H$. We do so via the following lemma:
\\~\\
\begin{lemma}
\label{lem:nu-bound-1}
If $S_1,~S_2$ and $\epsilon$ are defined as in Lemma~\ref{lem_main:search-highest-estate}, then
$$\sqrt{\Delta(1-\epsilon)}\leq \frac{S_1}{\sqrt{S_2}} \leq 1.$$
\end{lemma}
~\\~\\
\begin{proof}
The lower bound is obtained in a straightforward manner. Observe that
\begin{align}
\dfrac{S_1}{\sqrt{S_2}}&=\frac{\sum_{i<n}\frac{|a_{i}|^{2}}{1-\lambda_{i}}}{\sqrt{\sum_{i<n}\frac{|a_{i}|^{2}}{\left(1-\lambda{}_{i}\right)^{2}}}}\\
   &\geq \sqrt{\Delta\sum_{i<n}\frac{|a_{i}|^{2}}{1-\lambda_{i}}}\geq\sqrt{\Delta(1-\epsilon)}=\Theta\left(\sqrt{\Delta}\right),
\end{align}
It is possible to show that this bound is in fact tight. For example, we can construct a normalized Hamiltonian for which $|a_i|^2=1/n$ \footnote{This is possible independently of the position of the marked node $\ket{w}$, for example, if the Hamiltonian is diagonalized by the Fourier transform.} and the spectrum is such that:
~\\
\begin{itemize}
    \item there is one eigenvector with eigenvalue $1$.~\\ 
    \item there are $\Theta(n\sqrt{\Delta})$ eigenvectors with eigenvalue $1-\Delta$.~\\
    \item there are $\Theta(n(1-\sqrt{\Delta}))$ eigenvectors with eigenvalue $0$. 
\end{itemize}
~\\
It can be seen in this case the quantity $S_1/{\sqrt{S_2}}=\Theta(\sqrt{\Delta})$.

To prove the upper bound, we show that $S_1^2/S_2 \leq 1$, for which it suffices to prove that
$$S_2-S_1^2=\sum_{i<n}\dfrac{|a_{i}|^{2}}{(1-\lambda_{i})^2}-\left(\sum_{i<n}\dfrac{|a_{i}|^{2}}{1-\lambda_{i}}\right)^2> 0.$$
The left hand side of this inequality can be written as
\begin{align*}
\label{eq:sum}
&|a_n|^2\sum_{i<n}\dfrac{|a_{i}|^{2}}{(1-\lambda_{i})^2}+\sum_{i,k < n}\left(\frac{|a_{i}|^{2}|a_k|^2}{(1-\lambda_{i})^2}-\frac{|a_{i}|^{2}|a_k|^2}{(1-\lambda_{i})(1-\lambda_k)}\right).
\end{align*}
Clearly, the first term is always non-negative so we now show that the second term is also non-negative. The second term can be written as
\begin{align*}
&\sum_{i,k < n}\left(\frac{|a_{i}|^{2}|a_k|^2}{(1-\lambda_{i})^2}-\frac{|a_{i}|^{2}|a_k|^2}{(1-\lambda_{i})(1-\lambda_k)}\right)=\\
&\dfrac{1}{2}\sum_{i,k < n}\left(\frac{|a_{i}|^{2}|a_k|^2}{(1-\lambda_{i})^2}-\frac{|a_{i}|^{2}|a_k|^2}{(1-\lambda_{i})(1-\lambda_k)}\right)+\\
&\dfrac{1}{2}\sum_{k,i < n}\left(\frac{|a_{i}|^{2}|a_k|^2}{(1-\lambda_{k})^2}-\frac{|a_{i}|^{2}|a_k|^2}{(1-\lambda_{i})(1-\lambda_k)}\right)=\\
&\sum_{i,k < n}\dfrac{|a_i|^2|a_k|^2}{2}\left(\dfrac{1}{(1-\lambda_i)^2}+\dfrac{1}{(1-\lambda_k)^2}-\dfrac{2}{(1-\lambda_i)(1-\lambda_k)}\right)\\
&=\dfrac{1}{2}\left[\sum_{i,k < n}|a_i|^2|a_k|^2 \dfrac{(\lambda_k-\lambda_i)^2}{(1-\lambda_i)^2(1-\lambda_k)^2}\right]\geq 0.
\end{align*}
This implies that $S_1^2/S_2\leq 1$ and hence $S_1/\sqrt{S_2}\leq 1$. This is saturated (up to $O(1/n)$ terms), for example, if $H$ is the normalized adjacency matrix of the complete graph.
\end{proof}
\\~\\
Furthermore, we need to understand the validity of the spectral condition imposed in Eq.~\eqref{eq:spectral_con}. For this, it will be useful to write a weaker condition in terms of the spectral gap $\Delta$. From the definition of the quantities $S_k$ (see Eq.~\eqref{eqmain:Sk}) it is possible to see that $S_2\geq \Delta S_3$ and $S_1, S_2\geq 1$. Furthermore, from Lemma~\ref{lem:nu-bound-1} we have that $S_1\leq \sqrt{S_2}$. Hence, we can bound the RHS of the spectral condition as  
\begin{equation}\label{eq:spectral_con_bound}
 c \min\left\{\frac{S_1 S_2}{S_3},\Delta\sqrt{S_2}\right\}\geq c \Delta S_1  \geq c\Delta.
\end{equation}
This implies that our analysis is valid for any graph with $\sqrt{\epsilon}\leq c \Delta$ i.e., with a sufficiently large spectral gap compared to the overlap of the initial state with the marked node. For example, for $d$-dimensional lattices the spectral gap is $\Delta\sim n^{-2/d}$ and $\epsilon=1/n$ and so it is easy to see from the bound in Eq.~\eqref{eq:spectral_con} and \eqref{eq:spectral_con_bound} that the spectral condition is satisfied for lattices of dimension larger than $5$. In this scenario, we have that both $S_1$ and $S_2$ are constants \cite{childs2004spatial} and so we recover the result that marked node can be found in $\Theta(\sqrt{n})$ time in such a case as demonstrated by Childs and Goldstone. A similar behaviour appears in certain fractal lattices. The scaling of the gap depends on the spectral dimension $d_s$ as $\Delta\sim n^{-2/d_s}$ and the coefficients $S_1$ and $S_2$ are constant for spectral dimension larger than 4 \cite{Boettcher_evaluation_2017}. Hence, it can be shown that quantum search is optimal in this regime \cite{Boettcher_searchfractals_2017}. 

We nottice, in addition, that for regular lattices the performance of quantum search for the critical case $d=4$ is also recovered. For $4d$-lattices, we have that $\Delta=\Theta(1/\sqrt{n})$, $S_1=\Theta(1)$, $S_2=\Theta(\log n)$ and $S_3=\Theta(\sqrt{n})$ \cite{childs2004spatial}. As such the spectral condition is satisfied. 
Thus, the amplitude of the final state with the solution node is $S_1/\sqrt{S_2}=\Theta(1/\sqrt{\log n})$ after a time $T=\Theta(\sqrt{n\log n})$. It can be seen, however, than for dimensions $2$ and $3$ where the algorithm has been shown to be suboptimal \cite{childs2004spatial}, the spectral condition is violated and our analysis fails.

On the other hand, there exist graphs for which the $\C\G$ algorithm can be demonstrated to run in $\Theta(\sqrt{n})$ time, even though the spectral condition is violated. In the next section, we show how a different analysis helps capture the algorithmic performance on such instances. 

\section{Quantum search on graphs with quasi-degenerate highest eigenvalues}
\label{sec:search-degenerate-spectra}
In this section, we begin by considering examples of graphs for which the $\C\G$ algorithm runs optimally even though their normalized adjacency matrix violates the \textit{spectral condition} in Eq.~\eqref{eq:spectral_con}. 

\begin{figure}[t]
        \centering
        \begin{subfigure}{0.5\textwidth}
                \includegraphics[width=0.6\textwidth]{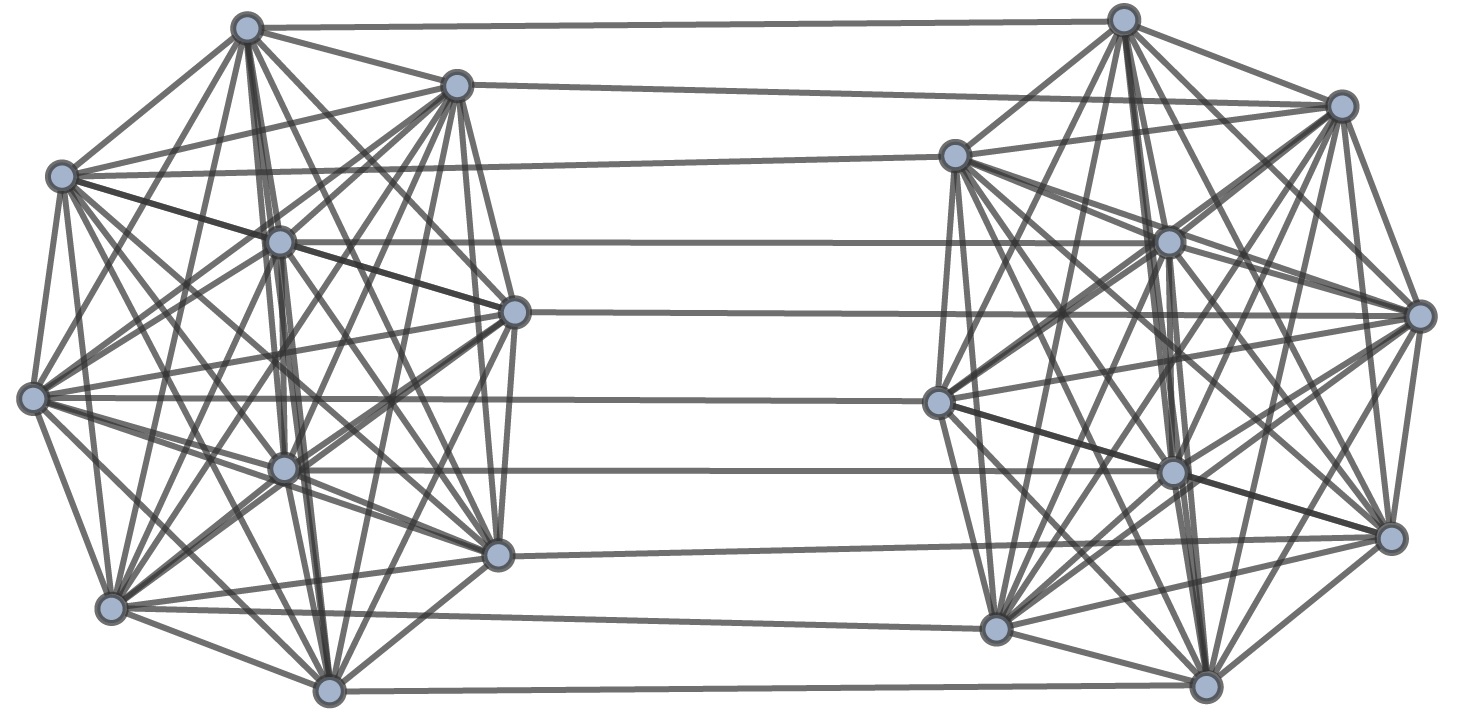}\\
                \vspace{2pt}
                \caption{~}
 				\label{subfig:bridged-complete}               
                \end{subfigure}\\
                
                \begin{subfigure}{0.5\textwidth}
                  \includegraphics[width=0.6\textwidth]{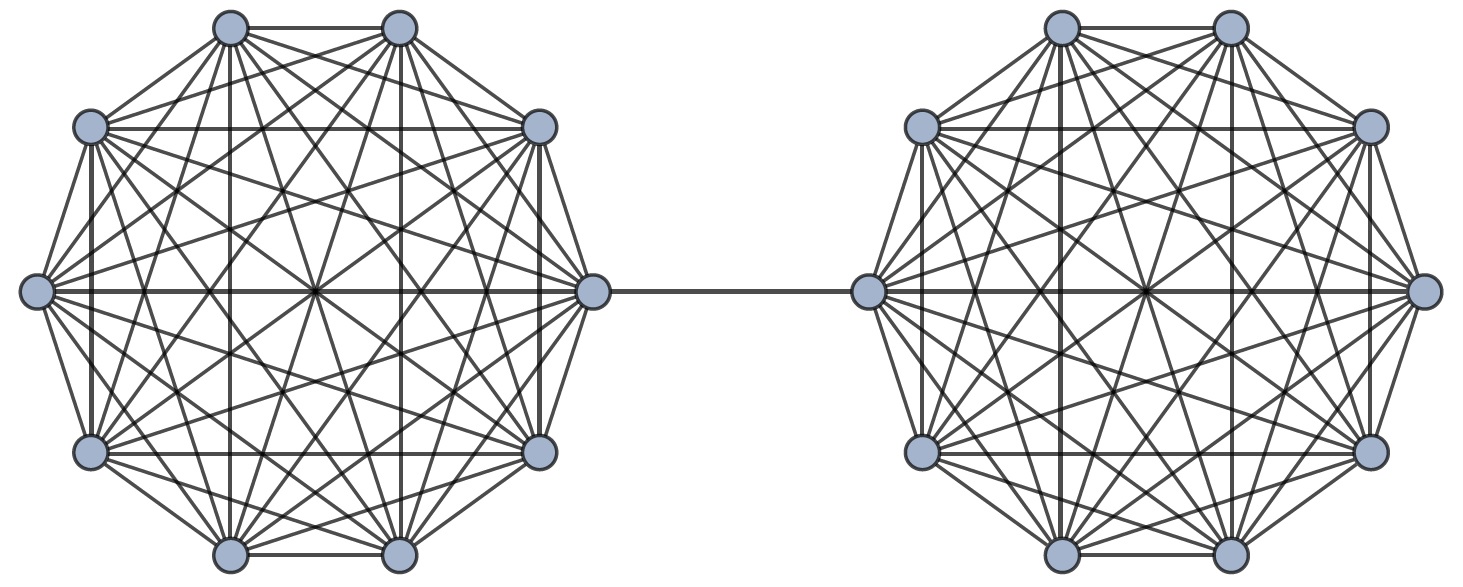}          
                \caption{~}
                \label{subfig:joined-complete}
                \end{subfigure}%
                \caption{\small{(a) The \textit{bridged-complete}} graph is a special case of a Rook's graph with $n_1=2$ and $n_2=n/2$. This corresponds to two complete graphs of $n/2$ edges such that each node in one complete graph is connected to the corresponding node in the other. (b) \textit{Joined-complete graph}: two complete graphs of $n/2$ nodes each are connected by a single edge.}
\label{figmain:bridged-joined-cg}
\end{figure}

An example of this is the following vertex-transitive graph, which we shall refer to as a \textit{bridged-complete graph}: two complete graphs of $n/2$ nodes such that every node in one complete graph is connected to the corresponding node in the other (See Fig.~\ref{subfig:bridged-complete}). This is a particular case of the Rook's graph which we discuss in  Sec.~\ref{sec:quantum-walk-chessboard}. The normalized adjacency matrix of this graph (i) has an extremely small spectral gap ($\Delta=\Theta\left(n^{-1+o(1)}\right)$) and (ii) there exists a constant gap between the first two eigenvalues and the rest of the spectrum. The spectral condition is violated, since
\begin{equation}
\min\left\{\frac{S_1 S_2}{S_3},\Delta\sqrt{S_2}\right\}=\Theta\left(\epsilon\right)\ll \sqrt{\epsilon},
\end{equation}
implying that Theorem \ref{lem_main:search-highest-estate} is not applicable for analyzing the algorithmic performance for this graph. 

However, intuitively the quantum walk search algorithm should run optimally for the bridged-complete graph. The quantum walk starts with an equal superposition of all nodes and, if we neglect the effect of the bridges connecting the two complete graphs, it is expected to be able to find a node marked in any of the two complete graphs of $n/2$ nodes with probability $1/2$ in $\sim \sqrt{n}/2$ time. Moreover, since there is a bridge connecting each node in one complete graph to another node in the other, the walker can transition between any of the two complete graphs. So one expects that a marked node would be obtained in $\Theta(\sqrt{n})$ time. A very similar example, with analogous spectral properties is the \textit{joined-complete graph} (two complete-graphs joined by a single bridge, Fig~\ref{subfig:joined-complete}). This example was used in Ref.~\cite{meyer2015connectivity} to show that large spectral gaps are indeed not necessary for optimal quantum search.

\begin{figure}
    \centering
    \includegraphics[scale=0.6]{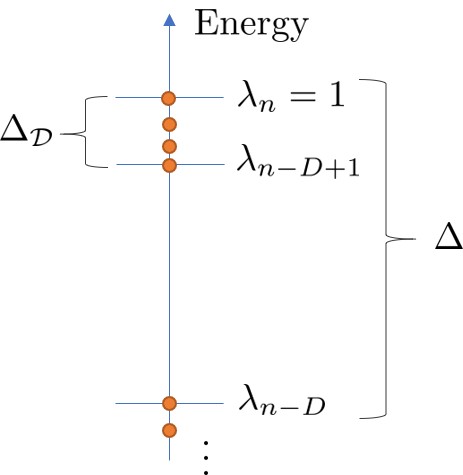}
    \caption{The analysis of Sec.~\ref{sec:search-degenerate-spectra} is tailored to the study of quantum search on graphs whose spectrum exhibits the features displayed in this figure. A small number $D$ of quasi-degenerate eigenvalues lie close to the maximum value $1$, within an energy gap $\Delta_{\De}$. The next largest eigenvalue has an energy $1-\Delta$, where $\Delta\gg \Delta_{\De}$. Such spectral features appear, for example, in the graphs of Fig~\ref{figmain:bridged-joined-cg} or, more generally, in graphs composed by highly connected clusters that are sparsely connected to each other.} 
    \label{fig:qd_eigenvalues}
\end{figure}

Thus, for both these graphs, we find that the spectrum of their normalized adjacency matrix satisfies the following two properties: (i)~A few of the highest eigenvalues are closely spaced (nearly degenerate) and (ii) there exists a large gap between these highest eigenvalues and the rest of the spectrum (see Fig.~\ref{fig:qd_eigenvalues}). We call the space spanned by the eigenvectors corresponding to the closely-spaced eigenvalues as \textit{quasi-degenerate}. Generally such graphs find applications in spectral clustering as they can be partitioned into clusters \cite{von2007tutorial}.

We show here that a modification of the analysis done in Sec.~\ref{sec:performance_non_degenerate}, which explicitly takes into account this quasi-degeneracy of the highest eigenvalues, allows us to construct spectral conditions which are sufficient to predict whether the $\C\G$ algorithm is optimal for Hamiltonians that satisfy the aforementioned properties. In particular, these conditions will allow us to predict optimality of quantum search for graphs such as the joined complete, or the bridged complete graph. 

Formally, consider a Hamiltonian $H$ such that its eigenvalues are
$$
0\leq\lambda_1\leq\cdots\leq \lambda_n=1,
$$ 
such that
$$
H\ket{v_i}=\lambda_i\ket{v_i}.
$$
Let us denote by $\De$ the space spanned by the $D$ eigenstates corresponding to the highest eigenvalues of $H$, i.e.\ 
\begin{equation}
\label{eq:quasi-degenerate-space}
\De=\mathrm{Span}\{\ket{v_n},\cdots,\ket{v_{n-D+1}}\},
\end{equation} 
such that $|\De|=D$. Consequently, we refer to the space spanned by the remaining eigenstates by $\bar{\De}$. Furthermore, we denote the gaps between the eigenvalues $\lambda_n=1$ and $\lambda_{n-D+1}$ as
\begin{equation}
\label{eqmain:gap-degenerate-space}
\Delta_\De= 1-\lambda_{n-D+1},
\end{equation}
and the gap between the $\lambda_n=1$ and $\lambda_{n-D}$ as 
\begin{equation}
\label{eqmain:gap-bet-nonD-D}
\Delta=1-\lambda_{n-D},
\end{equation}
as depicted in Fig.~\ref{fig:qd_eigenvalues}. Our analysis aims at predicting the algorithmic performance in cases where $\Delta_\De\ll \sqrt{\epsilon}$ and $\Delta $ is sufficiently large, for example, when $\Delta\gg \epsilon$. Hence, we say that the $D$ largest eigenvalues are nearly degenerate or \textit{quasi-degenerate}. The precise spectral properties that the Hamiltonian $H$ must fulfil are stated precisely in terms of a new spectral condition later on. Also, note that $D=1$ corresponds to the non-degenerate case considered in Theorem~\ref{lem_main:search-highest-estate} with $\Delta$ being the spectral gap of $H$. 

We demonstrate the algorithmic performance for such instances in the following subsections in two steps. We first assume that $\De$ is completely degenerate, i.e.\ all eigenstates in $\De$ have eigenvalue one ($\Delta_\De=0$) and find the evolution time and final amplitude of the algorithm based on this assumption (Subsec.~\ref{subsec:search-degenerate}) \footnote{From the Perron-Frobenius theorem, $\Delta_\De$ can never zero for the adjacency matrix (or Laplacian) of a connected graph.}. Next, we demonstrate that, given certain conditions on $\Delta_\De$, the algorithmic dynamics for the aforementioned case is the same as when $\De$ is  completely degenerate, up to some small error. 


\subsection{Performance of the $\mathbf{\C\G}$ algorithm when $\mathbf{\De}$ is degenerate}
\label{subsec:search-degenerate}
In order to analyse graphs with a $D$-degenerate highest eigenvalue ($\Delta_\De=0$), it will be useful to define the following quantities 
\begin{equation}
\label{eqmain:Sk-D}
S_{k,\bar{\De}}=\sum_{i=1}^{n-D}\dfrac{|a_i|^2}{(1-\lambda_i)^k},
\end{equation}
where $k\geq 1$. These parameters are similar to $S_k$ defined in Eq.~\eqref{eqmain:Sk}, except that the sum excludes all the $D$ degenerate eigenvalues (note that the quantities $S_k$ are not defined if there is degeneracy of the largest eigenvalue). 

Furthermore, we define $\sqrt{\epsilon_\De}$ as the overlap of the solution with the $\De$ subspace.  If the solution state $\ket{w}$ is expressed in the eigenbasis of $H$ as in Eq.~\eqref{eqmain:sol-in-eigenbasis}, this is given by 
\begin{equation}
\label{eqmain:overlap:w-D}
\epsilon_\De=\sum_{i\in \De} |a_i|^2.
\end{equation}
In addition, we define $\sqrt{\epsilon}=|\braket{w}{v_n}|=|a_n|$ as before, except that in this case $\ket{v_n}$ can be any state in the degenerate subspace $\De$. We introduce the following spectral condition, analogous to the one in Eq.~\eqref{eq:spectral_con}, in terms of the $S_{k,\bar{\De}}$ parameters
\begin{equation}
\label{eq:spec-cond-deg}
\sqrt{\epsilon_{\De}}\leq c\min\left\{\dfrac{S_{1,\bar{\De}}S_{2,\bar{\De}}}{S_{3,\bar{\De}}},\Delta\sqrt{S_{2,\bar{\De}}}\right\}.
\end{equation}
Our result regarding the performance of the quantum search algorithm is the following. 
~\\
\begin{theorem}
\label{thm:performance-search-deg}
Let $H$ be such that its largest $D$ highest eigenvalues are degenerate and the spectral condition of Eq.~\eqref{eq:spec-cond-deg} is obeyed, with $S_{k,\bar{\De}}$ defined as in Eq.~\eqref{eqmain:Sk-D}. By choosing $r=S_{1,\bar{\De}}$ and 
$$T=\Theta\left(\dfrac{1}{\sqrt{\epsilon_{\De}}}\dfrac{\sqrt{S_{2,\bar{\De}}}}{S_{1,\bar{\De}}}\right),$$
Algorithm \ref{algo-cg-general}, starting from any one of the $1$-eigenstates of $H$, denoted by $\ket{v_n}$, prepares a state $\ket{f}$ such that 
\begin{equation}\label{eq:nu_deg}
\nu=|\braket{w}{f}|=\Theta\left(\sqrt{\frac{\epsilon}{\epsilon_{\De}}}\frac{S_{1,\bar{\De}}}{\sqrt{S_{2,\bar{\De}}}}\right).
\end{equation}
\end{theorem}
~\\
\begin{proof}
Let the solution state $\ket{w}$ be expressed in the eigenbasis of $H$ as in Eq.~\eqref{eqmain:sol-in-eigenbasis}. It will be convenient to consider a rotated basis for the degenerate subspace $\De$ such that a single eigenstate, defined as 
\begin{equation}
\label{eqmain:state-relabeled-basis}
\ket{v_\De^{(1)}}=\dfrac{1}{\sqrt{\epsilon_\De}}\sum_{j\in\De} a_j\ket{v_j},
\end{equation}
contains the whole overlap of this subspace with $\ket{w}$, i.e. $|\braket{w}{v_\De^{(1)}}|=\sqrt{\epsilon_{\De}}$. We complete the basis with the states $\ket{v_\De^{(j)}}$, $2\leq j\leq D$, such that  $\braket{v_\De^{(l)}}{v_\De^{(m)}}=\delta_{l,m}$ for any $l,m\in \{1,...,D\}$. 

We note that this choice guarantees that $\braket{w}{v_\De^{(j)}}=0$, for $2\leq j\leq D$. This implies that these are eigenstates of $H_{\text{search}}$ with eigenvalue $1$, since
\begin{equation}\label{eq:trivial_eigenstates}
    H_{\text{search}}\ket{v_\De^{(j)}}= H \ket{v_\De^{(j)}}=\ket{v_\De^{(j)}}, 
\end{equation}
for $2\leq j \leq D$. This gives us a set of $D-1$ eigenstates which do not play a role in the computation of the final amplitude. To see this, let us first express the initial state $\ket{v_n}$ \footnote{Since we are assuming $D$ eigenstates with eigenvalue $1$, $\ket{v_n}$ can be any state in the subspace $\De$.} as
\begin{equation}
\label{eq:start-state-relabelled}
\ket{v_n}=\sum_{j=1}^{D}\alpha_j\ket{v_{\De}^{(j)}},
\end{equation} 
where $\sum_{j=1}^D |\alpha_j|^2=1$ and $\alpha_1=\sqrt{\epsilon/\epsilon_{\De}}$. We can now write
\begin{equation}\label{eq:amplitude_deg}
    \bra{w}e^{i H_{\text{search}}T} \ket{v_n} = \sqrt{\epsilon/\epsilon_{\De}} \bra{w}e^{i H_{\text{search}}T} \ket{v_\De^{(1)}},
\end{equation}
using Eq.~\eqref{eq:trivial_eigenstates} and the fact that $\braket{w}{v_\De^{(j)}}=0$, for $2\leq j\leq D$. 

Hence, to analyse the amplitude $\bra{w}e^{i H_{\text{search}}T} \ket{v_\De^{(1)}}$ it is enough to consider the dynamics in the subspace
\begin{equation}
    V=\text{span}\left\{\ket{v_\De^{(1)}}, \ket{v_i}, i\in\{1,...,n-D\}\right\}.
\end{equation}
We do this by applying the same techniques of Sec.~\ref{sec:performance_non_degenerate} for the projected search Hamiltonian
\begin{equation}
    H'_{\text{search}}=P_V H_{\text{search}} P_V =\ket{w}\bra{w} + P_V H P_V,
\end{equation}
where $P_V$ is the projector in the $V$ subspace. Note that the Hamiltonian $H'=P_V H P_V$ has a single eigenvalue 1 (the state $\ket{v_\De^{(1)}}$) and a spectral gap $\Delta$. The only difference with respect to the analysis in Sec.~\ref{sec:performance_non_degenerate} is that its dimension is $n-D+1$. Hence, we can apply Theorem~\ref{lem_main:search-highest-estate} by replacing the parameters $S_k$ by $S_{k,\bar{\De}}$ defined in Eq.~\eqref{eqmain:Sk-D} as well as replacing the spectral condition of Eq.~\eqref{eq:spectral_con} by the one in Eq.~\eqref{eq:spec-cond-deg}. This implies that, by choosing $r=S_{1,\bar{\De}}$ and evolving $H'_{\mathrm{search}}$ for time 
\begin{equation}\label{eq:tsearch_deg}
T=\Theta\left(\dfrac{\sqrt{S_{2,\bar{\De}}}}{S_{1,\bar{\De}}}\dfrac{1}{\sqrt{\epsilon_{\De}}}\right),
\end{equation}
with initial state $\ket{v_\De^{(1)}}$, results in a state that has an overlap with the solution 
\begin{equation}
\label{eqmain:amplitude-degenerate}
|\bra{w}e^{i H_{\text{search}}T} \ket{v_\De^{(1)}}|= \Theta\left(\frac{S_{1,\bar{\De}}}{\sqrt{S_{2,\bar{\De}}}}\right).
\end{equation}
Finally, replacing this amplitude in Eq.~\eqref{eq:amplitude_deg} we obtain that after this time the amplitude $| \bra{w}e^{i H_{\text{search}}T} \ket{v_n}|$ is given by Eq.~\eqref{eq:nu_deg}.
\end{proof}
~\\
 One can easily see that for $D=1$, $|\epsilon|=|\epsilon_{\De}|$, $S_{1,\bar{\De}}=S_1$ and $S_{2,\bar{\De}}=S_2$ we recover the statement of Theorem~\ref{lem_main:search-highest-estate}. However, for $D>1$ there is, in general, no way to have $|\epsilon|=|\epsilon_{\De}|$ as this would assume we are able to prepare the state $\ket{v_\De^{(1)}}$ from Eq.~\eqref{eqmain:state-relabeled-basis}, which depends on $w$ via the overlaps $a_i$. Given this, a possible strategy would be to choose $\ket{v_n}$ as a random state in the degenerate subspace $\De$, in which case the expected value of $\epsilon/\epsilon_{\De}$ would be $1/D$. 
 
Similarly to Theorem~\ref{lem_main:robustness_r}, it can be shown that the same algorithmic performance is maintained by choosing $r$ such that $|r-S_{1,\bar{\De}}|\ll \epsilon_{\De}S_{2,\bar{\De}}$. An analogous derivation to that of Theorem~\ref{lem:sub-optimality-for-any-r} shows that any choice of $r$ such that  
\begin{equation}
\label{eq:robust-range-of-r-deg}
r \not\in \left[S_{1,\bar{\De}}-\Theta\left(\sqrt{S_{2,\bar{\De}}\epsilon_\De}\right), S_{1,\bar{\De}}+\Theta\left(\sqrt{S_{2,\bar{\De}}\epsilon_\De}\right) \right],
\end{equation}
leads to a maximum amplitude of 
\begin{equation}
o\left(\sqrt{\frac{\epsilon}{\epsilon_{\De}}}\frac{S_{1,\bar{\De}}}{\sqrt{S_{2,\bar{\De}}}}\right).
\end{equation}

This implies the following necessary and sufficient conditions for optimality for Hamiltonians with degenerate highest eigenvalues. Provided that the spectral condition in Eq.~\eqref{eq:spec-cond-deg} holds, we obtain that the algorithm is optimal, in the sense discussed in Sec.~\ref{sec:def_optimality}, \emph{if and only if}
\begin{equation}\label{eq:cond_optimality_deg}
\frac{S_{1,\bar{\De}}}{\sqrt{S_{2,\bar{\De}}}}=\Theta(1),
\end{equation}
and $D=\Theta(1)$, which ensures that an amplitude of  $\sqrt{\epsilon/\epsilon_{\De}}=\Theta(1)$  after a time $T=\Theta(1/\sqrt{\epsilon})$.
More generally, if $D$ is not constant and provided  Eq.~\eqref{eq:cond_optimality_deg} holds, we obtain an amplitude 
$$
\nu\sim \sqrt{\frac{\epsilon}{\epsilon_\De}},
$$
after a time 
$$
T=\Theta\left(\dfrac{1}{\sqrt{\epsilon_\De}}\right).
$$
Hence, from Eq.~\eqref{eqmain:search-time-with-amp-amp},  using $\sqrt{\epsilon_{\De}/\epsilon}$ rounds of quantum amplitude amplification would also result in finding the marked vertex in time
\begin{equation}\label{eq:cost_deg_nuequal1}
T_{\mathrm{search}}=\sqrt{\dfrac{\epsilon_\De}{\epsilon}}\mathcal{S} + \dfrac{\C_w}{\sqrt{\epsilon}}+\M,
\end{equation}
which is similar to the optimal performance except that there is a multiplicative overhead in the set-up cost of $\sqrt{\epsilon_\De/\epsilon}$.
\subsection{Performance of the $	\mathbf{\C\G}$ algorithm when $\mathbf{\De}$ is quasi-degenerate}
\label{subsec:search-quasi-degenerate}
In this subsection, we explicitly calculate an upper bound on the error of the predicted performance of the $\CG$ algorithm when $\Delta_\De$,  defined in Eq.~\eqref{eqmain:gap-degenerate-space}, is larger than $0$. In this case, we write $H$ in its spectral form as
\begin{equation}
\label{eq:spectral-form-quasi-deg-Ham}
H=\sum_{j\in\De} \ket{v_j}\bra{v_j}+\sum_{j\notin\De} \lambda_j\ket{v_j}\bra{v_j}+\sum_{j\in\De} (\lambda_j-1)\ket{v_j}\bra{v_j}.
\end{equation}

This in turn implies that the search Hamiltonian $H_{\mathrm{search}}$ can be split as
\begin{align}
\label{eq:spectral-form-quasi-deg-Ham-search}
H_{\mathrm{search}}&=H^{\mathrm{deg}}_{\mathrm{search}}+H_{\mathrm{err}},
\end{align}
where $H^{\mathrm{deg}}_{\mathrm{search}}$ corresponds to the search Hamiltonian assuming that eigenspace $\De$ of $H$ is exactly degenerate with all eigenvalues in $\De$ equal to $1$, i.e.\
\begin{equation}
\label{eq:spectral-form-deg-Ham-search}
H^{\mathrm{deg}}_{\mathrm{search}}=\ket{w}\bra{w}+r \left(\sum_{j\in\De} \ket{v_j}\bra{v_j}+\sum_{j\notin\De} \lambda_j\ket{v_j}\bra{v_j}\right)
\end{equation}
and
\begin{equation}\label{eq:Herr}
H_{\mathrm{err}}=r \sum_{j\in\De} (\lambda_j-1)\ket{v_j}\bra{v_j}.
\end{equation}
We can quantify the error caused by neglecting $H_{err}$ in the time evolution of $H_{\mathrm{search}}$  for time  $T$ via the Trotter formulas \cite{childs2019trotter}. This leads to the following Lemma.
~\\
\begin{lemma}\label{lem:trotter-error}
Let $H_{\mathrm{search}}=H^{\mathrm{deg}}_{\mathrm{search}}+H_{\mathrm{err}}$, with $H^{\mathrm{deg}}_{\mathrm{search}}$ and $H_{\mathrm{err}}$ defined in Eqs.~\eqref{eq:spectral-form-deg-Ham-search} and \eqref{eq:Herr}, respectively.  For any time $T\geq 1/\sqrt{\epsilon_D}$, we have that 
\begin{equation}
  \bra{w}e^{i H_{\mathrm{search}}T} \ket{v_n}= \bra{w}e^{iH^{\mathrm{deg}}_{\mathrm{search}}T} \ket{v_n}+\eta_{qd},   
\end{equation}
where 
\begin{equation}
    \eta_{qd}=O\left(r \Delta_\De\sqrt{\epsilon_{\De}}T^2\right).
\end{equation}
\end{lemma}
~\\
\begin{proof}
Using first order Trotter formula \cite{childs2019trotter} we have 
\begin{align}
e^{-iTH_{\mathrm{search}}}&=e^{-iT\left(H^{\mathrm{deg}}_{\mathrm{search}}+H_{\mathrm{err}}\right)}\\
\label{eqmain:trotter-error1}                          
                          &=e^{-iTH^{\mathrm{deg}}_{\mathrm{search}}}e^{-iTH_{\mathrm{err}}}+O\left(\nrm{[\ket{w}\bra{w},H_{\mathrm{err}}]}T^2\right)\\
                          &=e^{-iTH^{\mathrm{deg}}_{\mathrm{search}}}e^{-iTH_{\mathrm{err}}}+O\left(r\Delta_\De\sqrt{\epsilon_{\De}}T^2\right),
\end{align}
where we used the bound 
\begin{equation}
    \nrm{[\ket{w}\bra{w},H_{\mathrm{err}}]}=O(r\Delta_\De\sqrt{\epsilon_{\De}}),
\end{equation}
which can be demonstrated by using the fact that $H_{\mathrm{err}}$ has support only on the $\De$ subspace. 

Moreover, we have that 
\begin{align}
e^{-iTH^{\mathrm{deg}}_{\mathrm{search}}}e^{-iTH_{\mathrm{err}}}&=e^{-iTH^{\mathrm{deg}}_{\mathrm{search}}}\left(I-iTH_{\mathrm{err}}+\cdots\right)\\
\label{eqmain:truncating-exponential}                                                                
                                                                &=e^{-iTH^{\mathrm{deg}}_{\mathrm{search}}}+O\left(T\nrm{H_{\mathrm{err}}}\right)\\
                                                                &=e^{-iTH^{\mathrm{deg}}_{\mathrm{search}}}+O\left(r\Delta_\De T\right).
\end{align}
The error in the approximation can be bounded by   
combining Eq.~\eqref{eqmain:trotter-error1} and Eq.~\eqref{eqmain:truncating-exponential} as
\begin{align}
\label{eq:error-trotter}
e^{-iTH_{\mathrm{search}}}=
&e^{-iTH^{\mathrm{deg}}_{\mathrm{search}}}
+\eta_{qd},
\end{align}
with
\begin{align}
   \eta_{qd}&= O\left(r\Delta_\De \max\left\{\sqrt{\epsilon_{\De}}T^2,T\right\}\right)
   \\&=O\left(r \Delta_\De\sqrt{\epsilon_{\De}}T^2\right),
\end{align}
where in the last step we assumed $T\geq 1/\sqrt{\epsilon_D}$.
\end{proof}
~\\
Using this Lemma, we can now adapt Theorem~\ref{thm:performance-search-deg} to Hamiltonians with quasi-degenerate highest eigenvalues. To do so, we need to impose a condition on the spectrum of $H$, that guarantees that the error $\eta_{qd}$ in Lemma~\ref{lem:trotter-error} is small enough for the predictions of Theorem~\ref{thm:performance-search-deg} to be meaningful. 
It can be seen that this is possible if 
\begin{align}
 \sqrt{\epsilon}\geq \dfrac{1}{c_1}\dfrac{S^{3/2}_{2,\bar{\De}}\Delta_\De}{S^2_{1,\bar{\De}}}.\label{eq:lowerbound_epsilonD}
\end{align}
For a sufficiently small positive constant $c_1$. A simpler, but less tight form for this condition in terms of the gaps $\Delta_\De$ and $\Delta$ can be obtained by using the lower bound in Lemma~\eqref{lem:nu-bound-1} (which is valid also for $S_{1,\bar{\De}}/\sqrt{S_{2,\bar{\De}}}$). It can be shown that if 
\begin{equation}
  \sqrt{\epsilon} \geq \frac{\Delta_\De}{ c_1 \Delta^2}
\end{equation}
then Eq.~\eqref{eq:lowerbound_epsilonD} is satisfied. For graphs such as the joined or bridged complete graph, we can take $D=2$ in which case $\Delta_{\De}=\Theta(1/n)$ and $\Delta=\Theta(1)$, whereas $\sqrt{\epsilon}=\Theta(n^{-1/2})$, ensuring this condition is satisfied.   

Our general result for search on graphs for which $H$ has quasi-degenerate highest eigenvalues is the following.
~\\
\begin{theorem}\label{thm:performance_qd}
For a given Hamiltonian $H$, assume there is a positive integer $D$ such that the conditions in Eqs.~\eqref{eq:lowerbound_epsilonD} and \eqref{eq:spec-cond-deg} are true. Then, by choosing $r=S_{1,\bar{\De}}$ and 
$$T=\Theta\left(\dfrac{1}{\sqrt{\epsilon_{\De}}}\dfrac{\sqrt{S_{2,\bar{\De}}}}{S_{1,\bar{\De}}}\right),$$
Algorithm \ref{algo-cg-general} prepares a state $\ket{f}$ such that 
\begin{equation}\label{eq:nu_deg}
\nu=|\braket{w}{f}|=\Theta\left(\sqrt{\frac{\epsilon}{\epsilon_{\De}}}\frac{S_{1,\bar{\De}}}{\sqrt{S_{2,\bar{\De}}}}\right).
\end{equation} 
\end{theorem}
~\\
\begin{proof}
The proof follows by combining the result of Lemma~\ref{lem:trotter-error} with that of Theorem~\ref{thm:performance-search-deg}. After a time
$$
T=\Theta\left(\dfrac{\sqrt{S_{2,\bar{\De}}}}{S_{1,\bar{\De}}}\dfrac{1}{\sqrt{\epsilon_{\De}}}\right),
$$
we have the following bound for the error term 
\begin{align}
\eta_{qd}&=O\left(\dfrac{\Delta_\De S_{2,\bar{\De}}}{\sqrt{\epsilon_{\De}}S_{1,\bar{\De}}}\right)\\
& =c_1 O\left(\sqrt{\frac{\epsilon}{\epsilon_{\De}}}\frac{S_{1,\bar{\De}}}{\sqrt{S_{2,\bar{\De}}}}\right),
\label{eq:additive-error-trotter}
\end{align}
where in the second step we use the condition from Eq.~\eqref{eq:lowerbound_epsilonD}.
Hence, for a sufficient small value of constant $c_1$, the amplitude obtained at the solution after evolving for this time $T$ from Eq.~\eqref{eq:tsearch_deg} is given by 
\begin{align}
    \bra{w}e^{i H_{\text{search}}T} \ket{v_n}& = \bra{w}e^{i H_{\text{search}}^{\mathrm{deg}}T}  \ket{v_n} + \eta_{qd}\\&
    =\Theta\left(\sqrt{\frac{\epsilon}{\epsilon_{\De}}}\frac{S_{1,\bar{\De}}}{\sqrt{S_{2,\bar{\De}}}}\right),
\end{align}
where we used Eq.~\eqref{eq:additive-error-trotter} and  Theorem~\ref{thm:performance-search-deg}.
\end{proof}

This result leads to the following \emph{sufficient} condition for optimal quantum search: provided there is an integer $D=\Theta(1)$ such that Eqs.~\eqref{eq:lowerbound_epsilonD} and \eqref{eq:spec-cond-deg} are satisfied and $S_{1,\bar{\De}}/{\sqrt{S_{2,\bar{\De}}}}=\Theta(1)$, quantum search is optimal. This is the case, for example, for graphs in which there a constant $D$ such that $\Delta_{\De}=o(\sqrt{\epsilon})$ and a large gap $\Delta=\Theta(1)$, such as the bridged or joined-complete graph.

Note that, even though in the fully degenerate case in Sec.~\ref{subsec:search-degenerate} our analysis provided necessary and sufficient conditions for optimal quantum search, here we can only provide a sufficient condition because our analysis gives no guarantee that the choice $r=S_{1,\bar{\De}}$ gives the best algorithmic performance when there is quasi-degeneracy. This is because, the error term that we obtain by approximating the quasi-degenerate case with the fully degenerate case in Lemma~\ref{lem:trotter-error}, becomes too large for sufficiently large values of $r$ and $T$.

\section{Finding a marked node on the Rook's graph} 
\label{sec:quantum-walk-chessboard}
In this section, we discuss the performance of the $\C\G$ algorithm on the Rook's graph. The edges of this graph correspond to the possible movements of a rook on a rectangular chessboard with $n=n_1 n_2$ nodes, where $n_1$ is the number of rows and $n_2$ the number of columns. We assume, without loss of generality, that $n_2\geq n_1$ and take $n_1=n^{\sigma}$ and $n_2=n^{1-\sigma}$ where $0<\sigma<1/2$.  

The motivation for studying quantum search on this graph is that, depending on the choice of $\sigma$, the performance of the algorithm varies drastically. The analysis of Sec.~\ref{sec:performance_non_degenerate} can be applied in certain regimes, showing that for $1/3\leq \sigma \leq 1/2$ the algorithm is optimal, whereas for $1/4 \leq \sigma < 1/3$ the algorithm is suboptimal and also slower than the square root of the classical hitting time, which for this graph is $\Theta(n)$. Interestingly, this suboptimality result holds even when the spectral gap $\Delta_{RG}\gg \sqrt{\epsilon}$ showing that the latter condition is not sufficient for optimal quantum search. 

For sufficiently low values of $\sigma$, the analysis of Sec.~\ref{sec:performance_non_degenerate} breaks down and the quasi-degenerate treatment from Sec.~\ref{sec:search-degenerate-spectra} can be used to provide lower bounds on the amplitude that can be obtained at the marked node after a certain time. This allows us, for example, to demonstrate that if $n_1=\Theta(1)$ the algorithm is optimal. Our predictions regarding the algorithmic performance are summarized in Table~\ref{tab:compare} for different regimes of $\sigma$.   

\subsection{The Rook's graph and its spectrum}
We begin by introducing the Rook's graph and the associated Hamiltonian $H$ that drives the quantum walk.

Consider the movement of a rook on a rectangular chessboard of $n_1$ rows and $n_2$ columns. The position of the rook on the chessboard is defined by the tuple $(i_{\leftrightarrow},j_{\updownarrow})$, where $i_{\leftrightarrow}\in [n_2]$ and $j_{\updownarrow}\in [n_1]$. From any given position, the rook can move horizontally (left or right) to any of the available $n_2$ positions or it can move vertically (forward and backward) to any of the available $n_1$ positions. Furthermore, suppose the rook accesses one of these available positions uniformly at random.
\begin{figure}[t]
        \centering
        \begin{subfigure}[b]{0.5\textwidth}
                \includegraphics[scale=0.4]{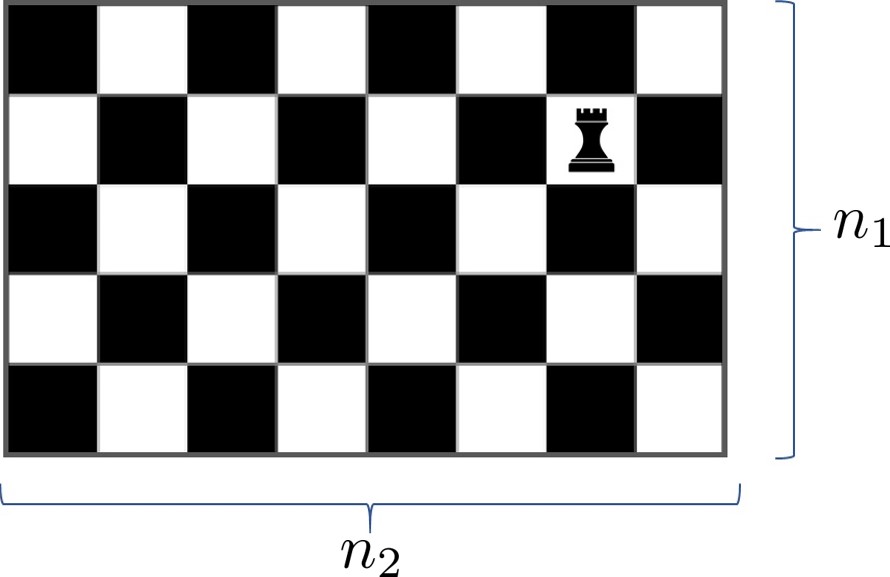}         
                \caption{~}
                \label{subfig:chessboard}
                \end{subfigure}
                
                \begin{subfigure}[b]{0.5\textwidth}
                \includegraphics[scale=0.45]{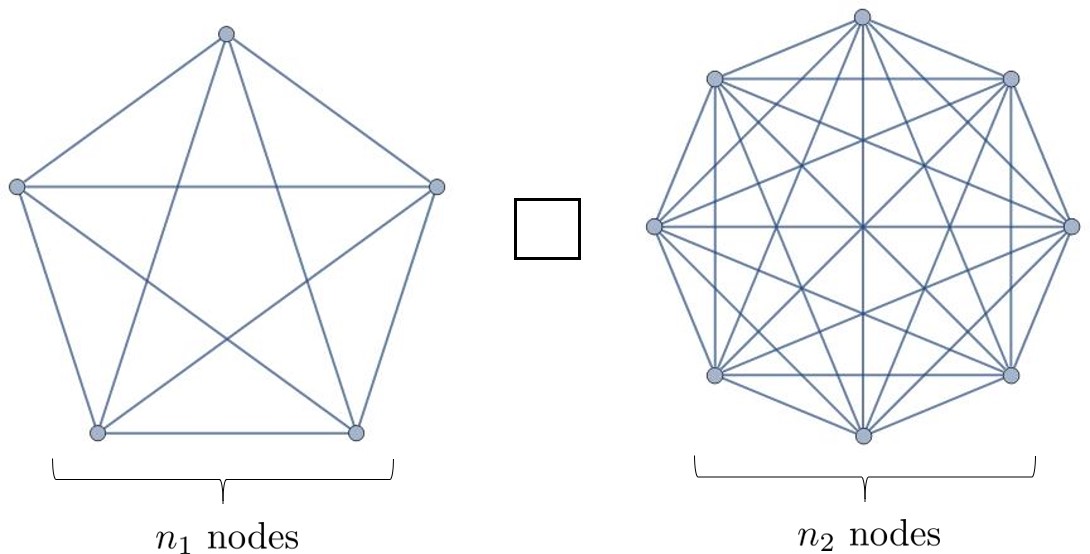}                
                \caption{~}
                \label{subfig:cartesian-complete-graphs}
                \end{subfigure}%
\caption{ The possible moves of a rook on a rectangular chessboard of $n_1$ rows and $n_2$ columns (a) corresponds to a graph which is the  Cartesian product of two complete graphs (Rook's graph) of $n_1$ and $n_2$ nodes respectively as depicted in (b).}
\label{figmain:chessboard}               
\end{figure}
If every cell of the chessboard is represented by the node of a graph then, the vertical movement of the rook is a walk on a complete graph of $n_1$ nodes and the horizontal movement corresponds to a walk on the complete graph of $n_2$ nodes. So, overall there are $n_2-1$ number of cliques (complete subgraphs) of $n_1$ nodes such that each node of an $n_1$-sized clique is connected to the corresponding node in $n_2-1$ other cliques. The resulting graph has $n=n_1n_2$ vertices and each node has degree $d=n_1+n_2-2$.  This regular graph, known as the \textit{Rook's graph} \cite{moon1963line,hoffman1964line}, corresponds to the Cartesian product of two complete graphs of $n_1$ nodes and $n_2$ nodes respectively and is also vertex-transitive. This has been depicted in Fig.~\ref{figmain:chessboard}. 

The Cartesian product of two graphs $G_1$ and $G_2$ is denoted as $G_1\square G_2$. If the adjacency matrix of $G_1$ is $A_{G_1}$ and the adjacency matrix of $G_2$ is $A_{G_2}$, then 
\begin{equation}
A_{G_1\square G_2}= A_{G_1}\otimes I_{n_2}+ I_{n_1}\otimes A_{G_2},
\end{equation}
where $\otimes$ denotes the Kronecker product and $I_j$ denotes the identity matrix of dimension $j$. Thus, the adjacency matrix of the Rook's graph is given by
\begin{align}
\label{eq:adj-matrix-graph}
A_G&= A^{n_1}_{CG}\otimes I_{n_2}+ I_{n_1}\otimes A^{n_2}_{CG},
\end{align}  
where $A^{n_1}_{CG}$ and $A^{n_2}_{CG}$ are the adjacency matrices of the complete graph with $n_1$ vertices and $n_2$ vertices respectively. 
As an aside, note that the graph corresponding to the case where $n_1=2$ and $n_2=n/2$ is the \textit{bridged-complete graph} see Fig.~\ref{subfig:bridged-complete} and the case where $n_2=n$ and $n_1=1$ is the complete graph.

We divide $A_G$ by the degree of each node $(n_1+n_2-2)$ and rescale its eigenvalues so they lie between $0$ and $1$. So the Hamiltonian we consider for the spatial search problem is the rescaled and shifted version of $A_G$ which we denote by $H$.
Without loss of generality, we take $n_2\geq n_1$ (the case $n_1\geq n_2$ can be recovered simply by exchanging the labels $1$ and $2$ i.e. what we refer to as horizontal and vertical directions). Furthermore, we assume that $n_1=n^{\sigma}$ and $n_2=n^{1-\sigma}$ where $0<\sigma<1/2$.  

It can be demonstrated that the Hamiltonian has 4 distinct eigenvalues (except in the case $n_1=n_2$, when there are there only three) which are shown in Table \ref{tab:evalues-chessboard} along with its degeneracies. Its spectral gap is given by
\begin{equation}
\label{eqmain:spectral-gap-chessboard}
\Delta_{RG}=1-\lambda_B=\Theta\left(\dfrac{n_1}{n_2}\right)=\Theta\left(\dfrac{1}{n^{1-2\sigma}}\right).
\end{equation}
Hence, by changing the value of $\sigma$, both the gap as well as the degeneracy of the different eigenvalues changes. 

In what follows we shall analyse the problem of finding a marked node on this graph for different regimes of $\sigma$. In particular, we will highlight regimes of $\sigma$ where treating the first few eigenstates of $H$ as quasi-degenerate shall help in deducing the algorithmic running time.  
\begin{table}[t]
\begin{center}
\begin{tabular}{{@{}lll@{}}}
\hline
Eigenvalue & Degeneracy \\ \hline
\vspace{4pt}
$\lambda_A=1$~~ &~~ $1$\\
\vspace{4pt}
$\lambda_B=\frac{n_2-\frac{1}{n_2}}{n_1+n_2-\frac{1}{n_1}-\frac{1}{n_2}}=1-\Theta(n^{2\sigma-1})$~~ &~~ $n_1-1$\\
\vspace{4pt}
$\lambda_C=\frac{n_1-\frac{1}{n_1}}{n_1+n_2-\frac{1}{n_1}-\frac{1}{n_2}}=\Theta(n^{2\sigma-1})$~~ &~~ $n_2-1$\\
\vspace{4pt}
$\lambda_D=0$~~&~~ $(n_1-1)(n_2-1)$\\ 
\hline
\end{tabular}
\caption{Eigenvalues (along with their respective degeneracies) of the Hamiltonian $H$ corresponding to the normalized adjacency of the Rook's graph }
\label{tab:evalues-chessboard}
\end{center}
\end{table}
\subsection{Algorithmic performance when $\mathbf{r=S_1}$}
\label{subsec:r-S1-chessboard} 
We will first analyse the performance of quantum search via the approach developed in Sec.~\ref{sec:performance_non_degenerate}. In order to determine the regime of validity of this approach we need to estimate the quantities $\epsilon, S_1$, $S_2$ and $S_3$. First observe that the resultant graph is symmetric and vertex-transitive, i.e.\ $|a_i|=1/\sqrt{n},~\forall i$, where $a_i$ is as defined in Eq.~\eqref{eqmain:sol-in-eigenbasis}. Consequently, we have $\epsilon=1/n$. 

Furthermore, using the definition of the parameters $S_k$ from Eq.~\eqref{eqmain:Sk}, it can be shown that these parameters scale with $n$ as  
\begin{align}
    S_k&= \Theta\left(\frac{n^{\sigma-1}}{(n^{2\sigma-1})^k}+1\right).
\end{align}
In particular, this implies that 
\begin{align}
S_1=\Theta(1)
\end{align}
and
\begin{equation}
S _2 = \begin{cases}
        \Theta\left(n^{1-3\sigma}\right),~&\text{for } 0< \sigma\leq 1/3\\
        \Theta\left(1\right),~&\text{for } 1/3\leq \sigma \leq 1/2.
        \end{cases}    
\end{equation}
We can now verify the regime of validity of \textit{spectral condition} in Eq.\eqref{eq:spectral_con}, which is required for Theorem \ref{lem_main:search-highest-estate} to be applied. It is easy to verify that this holds only when $1/4\leq \sigma \leq 1/2$. 
Consequently the amplitude of the final state of the algorithm with the marked vertex is
\begin{align}
\nu=\Theta\left(\frac{S_1}{\sqrt{S_2}}\right)&=\begin{cases}
       \Theta\left(n^{-(1-3\sigma)/2}\right),~&\text{for } 1/4\leq \sigma\leq 1/3\\
       \Theta\left(1\right),~&\text{for } 1/3\leq \sigma \leq 1/2,
        \end{cases}
\end{align}
after a time 
\begin{align}
T=\Theta\left(\dfrac{1}{\sqrt{\epsilon}}\dfrac{\sqrt{S_2}}{S_1}\right)&=\begin{cases}
       \Theta\left(n^{1-3\sigma/2}\right),~&\text{for } 1/4\leq \sigma\leq 1/3\\
       \Theta\left(\sqrt{n}\right),~&\text{for } 1/3\leq \sigma \leq 1/2.
       \end{cases}
\end{align}
The performance of the algorithm is thus quite distinct in the following two regimes. 
\\~\\
\textbf{\textit{i) $\mathbf{1/3\leq \sigma\leq 1/2}$:}~} In this regime the algorithm is optimal, since $\nu=\Theta(1)$ after a time $\Theta(\sqrt{n})$. This provides another example of optimal search for graphs which do not have constant spectral gap. In fact, from Eq.~\eqref{eqmain:spectral-gap-chessboard} we see that in this regime the scaling of the spectral gap changes from $n^{-1/3}$ for $\sigma=1/3$ to constant for $\sigma=1/2$ .
\\~\\
\textbf{\textit{ii) $\mathbf{1/4\leq \sigma< 1/3}$:}~} In this regime, the algorithm is suboptimal, as the final overlap with the marked node $\nu=\Theta\left(n^{-(1-3\sigma)/2}\right)$ after $T= \Theta\left(n^{1-3\sigma/2}\right)$. In the worst case, for $\sigma=1/4$, even if we assume that amplitude amplification can be used, one would need to evolve the walk for a total time of $\Theta(n^{3/4})$ to find the marked node. 
~\\
Interestingly, the Rook's graph within this region of $\sigma$ provides an example of suboptimality even though in this regime we have that $\Delta_{RG}\gg \sqrt{\epsilon}=n^{-1/2}$ (excluding in the case $\sigma=1/4$). This proves that the latter condition is not sufficient for optimal quantum search. In addition, given that the hitting time for the Rook's graph is $\Theta(n)$ for any $\sigma$, this shows that there exists a range of values of $\sigma$ for which the $\C\G$ algorithm is slower than the square root of the classical hitting time.

\subsection{Algorithmic performance when $\mathbf{r=S_{1,\bar{\mathcal{\mathbf{D}}}}}$}
In order to go beyond the limitations imposed by the spectral condition of Eq.~\eqref{eq:spectral_con}, which is only valid in the regime $1/4\leq \sigma\leq 1/2$, we use the analysis of Sec.~\ref{sec:search-degenerate-spectra}. It is expected that this analysis is valid for low values of $\sigma$, since the gap $1-\lambda_B$ becomes very small (Eq.~\eqref{eqmain:spectral-gap-chessboard}) whereas the gap $1-\lambda_C=\Theta(1)$ is much larger (see Table~\ref{tab:evalues-chessboard}).  

Hence, we will treat the eigenstate with eigenvalue $\lambda_A=1$ and the $n_1-1$ eigenstates with eigenvalue $\lambda_B$  as \textit{quasi-degenerate}. To be consistent with the notation in Sec.~\ref{sec:search-degenerate-spectra}, we denote this space as $\De$ such that $D=|\De|=n_1=n^\sigma$. In addition, the gaps $\Delta_\De$ and $\Delta$ defined in Sec.~\ref{sec:search-degenerate-spectra} are in this case
\begin{align}
  \Delta_\De&\equiv \Delta_{RG}=1-\lambda_B=\Theta(n^{2\sigma-1}),\\
  \Delta&=1-\lambda_C=\Theta(1).
\end{align}
The projection of the marked node $\ket{w}$ in the quasi-degenerate subspace $\De$ is 
\begin{equation}
\label{eqmain:epsilon-D-chessboard}
\epsilon_\De=\frac{D}{n}=\frac{n_1}{n},
\end{equation}
where we have used the fact that the underlying graph is vertex-transitive implying that $|a_i|^2=1/n,~\forall i$.
The parameters $S_{k,\bar{\De}}$, defined in Eq.~\eqref{eqmain:Sk-D}, are given by
\begin{align}
S_{k,\bar{\De}}&=\frac{1}{n} \left[\dfrac{n_2-1}{(1-\lambda_C)^k}+\dfrac{(n_1-1)(n_2-1)}{(1-\lambda_D)^k}\right]\\
		 &=1+\Theta(1/n_1)=\Theta(1),
\end{align}
$\forall k\geq 1$.  

For Theorem \ref{thm:performance_qd} to hold, we require that both conditions Eqs.~\eqref{eq:lowerbound_epsilonD} and \eqref{eq:spec-cond-deg} are satisfied. Since $\Delta$ and the parameters $S_{k,\bar{\De}}$ are constants, the condition in Eq.~\eqref{eq:spec-cond-deg} is valid for any $0\leq \sigma < 1/2$. On the other hand, it can be shown that \eqref{eq:lowerbound_epsilonD} is valid as long as $\sigma\leq 1/4$. Hence, Theorem \ref{thm:performance_qd} allows us to predict the performance of the algorithm for the choice of $r=S_{1,\bar{\De}}$ and in the regime  $0\leq \sigma\leq 1/4$.

\renewcommand{\arraystretch}{2.2}
\begin{table*}[ht]
\begin{center}
\begin{tabular}{c|c|c|c|c}
\hline
\multirow{2}{*}{Range of $n_1=n^\sigma$} & \multicolumn{2}{c|}{$r=S_1$} & \multicolumn{2}{c}{$r=S_{1,\bar{\De}}$} \\ \cline{2-5} 
                              & $T$         & $\nu$         &      $T$          &     $\nu$         \\ \cline{1-5}
\cline{1-5}
           $n_1=\Theta(1)$          	  & -- & -- &$\Theta\left(\sqrt{n}\right)$& $\Theta(1)$          \\
           $0<\sigma<1/4$          &  -- & -- & $\Theta\left(\sqrt{\frac{n}{n^\sigma}}\right)$	&  $\Theta\left(\frac{1}{\sqrt{n^\sigma}}\right)$                  \\
            $1/4\leq \sigma <1/3$          &  $\Theta\left(n^{1-3\sigma/2}\right)$& $\Theta\left(n^{-(1-3\sigma)/2}\right)$& --	&  --\\
           	$1/3\leq \sigma\leq 1/2$             &    $\Theta\left(\sqrt{n}\right)$         &   $\Theta(1)$            &      --             &      --             \\
\hline
\end{tabular}
\caption{\small{Summary of the performance of the $\C\G$ algorithm (Algorithm~1) on the Rook's graph, corresponding to the movement of a rook on a rectangular chessboard of $n_1=n^{\sigma}$ columns and $n^{1-\sigma}$ rows for different regimes of $\sigma$ and for two different choices of the parameter $r$. When $0\leq \sigma < 1/4$, treating the highest eigenvalues of the Hamiltonian associated with the graph as quasi-degenerate allows us to analyse the algorithmic performance via Theorem \ref{thm:performance_qd} for the choice of $r=S_{1,\bar{\De}}$ and predict optimality when $n_1$ is constant. On the other hand, Theorem \ref{lem_main:search-highest-estate} is valid in the complementary regime of $1/4\leq \sigma\leq 1/2$, allowing us to predict the best algorithmic performance for these values of $\sigma$. We conclude that the algorithm is optimal when $1/3\leq \sigma\leq 1/2$ as a marked node can be found in $\sqrt{n}$ time. In contrast, in the regime $1/4\leq \sigma< 1/3$ the maximum amplitude at the marked node is $o(1)$ implying suboptimality (see also Theorem~\ref{thm:optimalityCG}). Finally, the results for $1/2\leq \sigma\leq 1$ can be obtained from the results of the regime $0\leq \sigma \leq 1/2$ by replacing $\sigma$ with $1-\sigma$}}
\label{tab:compare}
\end{center}
\end{table*}

We obtain that, after a time
\begin{equation}
\label{eq:time-chessboard}
T=\Theta\left(\dfrac{\sqrt{S_{2,\bar{\De}}}}{S_{1,\bar{\De}}}\dfrac{1}{\sqrt{\epsilon_{\De}}}\right)=\Theta\left(\sqrt{\dfrac{n}{n_1}}\right)=\Theta\left(\sqrt{n^{1-\sigma}}\right),
\end{equation}
Algorithm~\ref{algo-cg-general} prepares a state that has an overlap of 
\begin{equation}
\label{eq:amp-chessboard}
\nu =\Theta\left(\sqrt{\dfrac{\epsilon}{\epsilon_\De}}\frac{S_{1,\bar{\De}}}{\sqrt{S_{2,\bar{\De}}}}\right)=\Theta\left(\frac{1}{\sqrt{n_1}}\right)=\Theta\left(\frac{1}{\sqrt{n^\sigma}}\right)
\end{equation} 
with the marked node. We discuss the following two cases:
\\~\\
\textbf{\textit{i)~constant $\mathbf{n_1}$ $(\mathbf{\sigma=0})$:}~} In this regime, the algorithm is optimal, as the marked node is found with a constant probability after $T=\Theta\left(\sqrt{n}\right)$. Note that, as mentioned before, the \textit{bridged-complete graph} corresponds to the choice of $n_1=2$ and $n_2=n/2$. As such, this demonstrates the optimality of the algorithm for this instance. 
\\~\\
\textbf{\textit{ii)~$\mathbf{n_1=n^\sigma}$, with~$\mathbf{0<\sigma<1/4}$:}~} In this range, the amplitude at the marked node is $\Theta(1/\sqrt{n_1})$ after $\Theta(\sqrt{n/n_1})$ time. If the quantum amplitude amplification procedure is available, then using $\sqrt{n_1}$ rounds of amplitude amplification, the marked node can be obtained for $T=\Theta(\sqrt{n})$, however, with an overhead due to the need of reflecting on the initial state that has a certain set-up cost (in this case the total cost is given by Eq.~\eqref{eq:cost_deg_nuequal1}). 

On the other hand, if we assume access to simply the time evolution according to $H_{\mathrm{search}}$, we have to repeat the entire procedure $\sim n_1$ times to find the marked node, leading to an overall evolution time of $T=\Theta(\sqrt{n.n_1})$. However, as we previously pointed out, the predicition  from Theorem~\ref{thm:performance_qd} does not guarantee that this is the best possible performance. We leave open the question of whether a better running time can be obtained for a different choice of $r$. Indeed, one would expect that the best choice of $r$ should converge to the value $S_1$ at $c=1/4$ and recover the prediction of Sec.~\ref{subsec:r-S1-chessboard}.
\\~\\ 
\section{Discussion} 
\label{sec:discussion}
In this article, we provide the necessary and sufficient conditions for the $\C\G$ algorithm to be optimal, assuming very general conditions on the spectrum of the Hamiltonian encoding the structure of the underlying graph. An immediate consequence is that our necessary and sufficient conditions hold for all graphs whose normalized adjacency matrices exhibit a large enough spectral gap ($\Delta\gg \sqrt{\epsilon}$). Additionally, we also provide strategies to analyze the algorithmic performance for graphs with a few quasi-degenerate highest eigenvalues, followed by a large gap. Such spectral features appear, for example, in graphs composed by a few clusters with sparse connections among them. Our work implies that, to the best of our knowledge,  all prior results demonstrating the optimality of the algorithm for specific graphs, requiring instance-specific analysis, can now be recovered from our general results. We also provided an explicit example, namely, the application of the $\C\G$ algorithm to the Rook's graph which highlights the predictive power of our results and the limitations of this search algorithm, which is suboptimal for certain regimes of the ``aspect ratio'' of the chessboard.

Our results provide a recipe to compute analytically the performance of the $\C\G$ algorithm on any graph fulfilling the spectral conditions required for our main theorems to be valid (Theorem \ref{lem_main:search-highest-estate} and \ref{thm:performance_qd}). They can hence be used to analyse quantum search on graphs that have not been previously studied or on graphs that were analysed only numerically such as the Chimera graph~\cite{glos2019impact} -- a graph that encodes the underlying architecture of the hardware of quantum annealers.

We remark that our results are not directly applicable, but could in principle be extended, to some modified versions of the $\C\G$ algorithm. For example, it would be interesting to extend our analysis to encompass strategies with different choices of the parameter $r$ for different evolution times. Such strategies have been known to improve the algorithmic performance for some graphs \cite{meyer2015connectivity}. We also note that our results are only valid for the oracle Hamiltonian introduced in \cite{childs2004spatial}, which singles out the marked node by adding a local energy term to the Hamiltonian. Different oracles, which remove edges connected to the marked node, have been considered in works that show optimal quantum search on certain lattices such as graphene \cite{foulger2014quantum} or crystal lattices \cite{childs2014spatial-crystal}. General conditions for optimal quantum search with such oracles are still unknown (some progress has been made in \cite{chakraborty2018finding}).

Our work further highlights the difficulty in comparing the performance of the $\C\G$ algorithm to its classical counterpart (search by a classical random walk), where the performance is measured by the classical hitting time. In fact, it is not clear how the expressions that we have obtained for predicting the performance of this quantum search algorithm relate to this classical quantity. We can nevertheless guarantee that whenever the $\C\G$ algorithm is optimal, there exists at least a quadratic speed-up with respect to the classical hitting time, since the latter is lower bounded by $1/\epsilon$ (see Eq.~\eqref{eqmain:hitting-time-bounds}). However, we also proved that the $\C\G$ algorithm fails to achieve quadratic speedups with respect to classical search in some cases, as evidenced by the algorithmic running time on the Rook's graph in the suboptimal regime (See Eq.~\eqref{eq:time-chessboard} and Eq.~\eqref{eq:amp-chessboard}).

Different quantum walk based algorithms are known to achieve this general quadratic speed-up in the discrete-time framework \cite{szegedy2004quantum,krovi2016quantum,ambainis2019quadratic} and in continuous-time, as recently demonstrated in Ref.~\cite{chakraborty2018finding}. In the latter work, we propose a new continuous-time time quantum walk algorithm based on the time-evolution of a Hamiltonian encoding an interpolating Markov chain. Compared to the $\C\G$ algorithm, it has the additional advantages that it can be applied to any ergodic, reversible Markov chain and has a guaranteed performance even when there are multiple marked nodes. 

We also propose a modified version of the $\C\G$ algorithm that is applicable to search problems on Markov chains~\cite{chakraborty2018finding} which can be seen as a quantum walk on the edges and thus requires an extension of the walk's Hilbert space. This modified algorithm improves upon the performance of the original $\C\G$ algorithm for several instances such as lattices of dimension less than five. In particular, this modified algorithm can find a marked node on the Rook's graph in $\Theta(\sqrt{n})$ time for any dimensions of the chessboard, without requiring amplitude amplification. However, the modified $\C\G$ algorithm does not provide a generic speedup over the $\C\G$ algorithm and counterexamples have also been demonstrated in Ref.~\cite{chakraborty2018finding}. Moreover, a simple comparison of the performance of this modified $\C\G$ algorithm to the classical hitting time remains elusive.  

An interesting direction of future research would be to explore the possibility of using continuous-time quantum walks to solve optimization problems, namely, to find ground states of classical Ising Hamiltonians which encode the solution to some NP-Hard problems \cite{lucas2014ising}. Recently, the applicability of CTQW to tackle this problem has been numerically investigated \cite{callison2019finding}. In this approach, each node represents a spin configuration, and the idea is to perform a continuous-time quantum walk on a graph where the local energies of each node, instead of being set by an oracle, corresponds to the energy of this respective configuration according to the Ising Hamiltonian. The aim is thus to use the quantum walk to find the node of minimum energy faster than classical methods. In Ref~\cite{callison2019finding}, the authors observe that this approach leads a faster than quadratic speedup, with respect to unstructured search, in the time required to find the ground state of random Ising Hamiltonians, for quantum walks on certain graphs. We remark that when the underlying graph of this quantum walk is the complete graph, our results can be used to make some analytical predictions, since the Hamiltonian of the walk has the form $H+r P$, where $P$ is a one-dimensional projector (in this case, $H$ is the classical Ising Hamiltonian and the adjacency matrix of the complete graph is a one-dimensional projector). It would be interesting if extensions of our results could help derive analytical expressions for the performance of this approach on different graphs. This could lead to a better understanding of the potential of CTQW-based algorithms to solve optimization problems.
\\~\\
\begin{acknowledgments}
S.C. and J.R. are supported by the Belgian Fonds de la Recherche Scientifique - FNRS under grants no F.4515.16 (QUICTIME) and R.50.05.18.F (QuantAlgo). S.C. and L.N. acknowledge support from F.R.S.-FNRS. L.N. also acknowledges funding from the Wiener-Anspach Foundation.
\end{acknowledgments}
\bibliographystyle{unsrt}
\bibliography{References}

\begin{thebibliography}{10}

\bibitem{kempe2003quantum}
Julia Kempe.
\newblock Quantum random walks: an introductory overview.
\newblock {\em Contemporary Physics}, 44(4):307--327, 2003.

\bibitem{childs2009universal}
Andrew~M Childs.
\newblock Universal computation by quantum walk.
\newblock {\em Physical review letters}, 102(18):180501, 2009.

\bibitem{childs2013universal}
Andrew~M Childs, David Gosset, and Zak Webb.
\newblock Universal computation by multiparticle quantum walk.
\newblock {\em Science}, 339(6121):791--794, 2013.

\bibitem{ambainis2003quantum}
Andris Ambainis.
\newblock Quantum walks and their algorithmic applications.
\newblock {\em International Journal of Quantum Information}, 1(04):507--518,
  2003.

\bibitem{childs2004spatial}
Andrew~M Childs and Jeffrey Goldstone.
\newblock Spatial search by quantum walk.
\newblock {\em Physical Review A}, 70(2):022314, 2004.

\bibitem{Note1}
Throughout the article, we use a plethora of complexity theoretic notations
  which we briefly define now. $f(n)=\protect \mathcal {O}(g(n))$, if there
  exists a positive constant $c$ such that $|f(n)|\leq c. g(n)$. If
  $f(n)=\protect \mathcal {O}(g(n))$, then $g(n)=\Omega (f(n))$. Also
  $f(n)=o(g(n))$ if for all positive constants $k$, $|f(n)|<k.g(n)$.
  Consequently, if $f(n)=o(g(n))$, then $g(n)=\omega (f(n))$.

\bibitem{janmark2014global}
Jonatan Janmark, David~A Meyer, and Thomas~G Wong.
\newblock Global symmetry is unnecessary for fast quantum search.
\newblock {\em Physical Review Letters}, 112(21):210502, 2014.

\bibitem{novo2015systematic}
Leonardo Novo, Shantanav Chakraborty, Masoud Mohseni, Hartmut Neven, and Yasser
  Omar.
\newblock Systematic dimensionality reduction for quantum walks: Optimal
  spatial search and transport on non-regular graphs.
\newblock {\em Scientific reports}, 5:13304, 2015.

\bibitem{chakraborty2016spatial}
Shantanav Chakraborty, Leonardo Novo, Andris Ambainis, and Yasser Omar.
\newblock Spatial search by quantum walk is optimal for almost all graphs.
\newblock {\em Physical review letters}, 116(10):100501, 2016.

\bibitem{philipp2016continuous}
Pascal Philipp, Lu{\'\i}s Tarrataca, and Stefan Boettcher.
\newblock Continuous-time quantum search on balanced trees.
\newblock {\em Physical Review A}, 93(3):032305, 2016.

\bibitem{wong2016quantum}
Thomas~G Wong.
\newblock Quantum walk search on johnson graphs.
\newblock {\em Journal of Physics A: Mathematical and Theoretical},
  49(19):195303, 2016.

\bibitem{chakraborty2017optimal}
Shantanav Chakraborty, Leonardo Novo, Serena Di~Giorgio, and Yasser Omar.
\newblock Optimal quantum spatial search on random temporal networks.
\newblock {\em Physical review letters}, 119(22):220503, 2017.

\bibitem{Boettcher_searchfractals_2017}
Shanshan Li and Stefan Boettcher.
\newblock Renormalization group for a continuous-time quantum search in finite
  dimensions.
\newblock {\em Phys. Rev. A}, 95:032301, Mar 2017.

\bibitem{wong2018quantum}
Thomas~G Wong, Konstantin W{\"u}nscher, Joshua Lockhart, and Simone Severini.
\newblock Quantum walk search on kronecker graphs.
\newblock {\em Physical Review A}, 98(1):012338, 2018.

\bibitem{glos2018optimal}
Adam Glos and Thomas~G Wong.
\newblock Optimal quantum-walk search on kronecker graphs with dominant or
  fixed regular initiators.
\newblock {\em Physical Review A}, 98(6):062334, 2018.

\bibitem{rhodes2019quantum}
Mason~L Rhodes and Thomas~G Wong.
\newblock Quantum walk search on the complete bipartite graph.
\newblock {\em Physical Review A}, 99(3):032301, 2019.

\bibitem{Note2}
It is worth mentioning, that throughout the article we consider the scenario
  where only a single node is marked and analogous results for multiple marked
  nodes is an open problem in this framework.

\bibitem{farhi1998analog}
Edward Farhi and Sam Gutmann.
\newblock Analog analogue of a digital quantum computation.
\newblock {\em Physical Review A}, 57(4):2403, 1998.

\bibitem{meyer2015connectivity}
David~A Meyer and Thomas~G Wong.
\newblock Connectivity is a poor indicator of fast quantum search.
\newblock {\em Physical review letters}, 114(11):110503, 2015.

\bibitem{von2007tutorial}
Ulrike Von~Luxburg.
\newblock A tutorial on spectral clustering.
\newblock {\em Statistics and computing}, 17(4):395--416, 2007.

\bibitem{moon1963line}
JW~Moon.
\newblock On the line-graph of the complete bigraph.
\newblock {\em The Annals of Mathematical Statistics}, pages 664--667, 1963.

\bibitem{hoffman1964line}
AJ~Hoffman et~al.
\newblock On the line graph of the complete bipartite graph.
\newblock {\em The Annals of Mathematical Statistics}, 35(2):883--885, 1964.

\bibitem{Note3}
This can be ensured by replacing $H$ with $(H/||H||+I)/2$, where $I$ is the
  identity matrix.

\bibitem{roland2003quantum}
J{\'e}r{\'e}mie Roland and Nicolas~J Cerf.
\newblock Quantum-circuit model of hamiltonian search algorithms.
\newblock {\em Physical Review A}, 68(6):062311, 2003.

\bibitem{bollobas2013modern}
B{\'e}la Bollob{\'a}s.
\newblock {\em Modern graph theory}, volume 184.
\newblock Springer Science \& Business Media, 2013.

\bibitem{childs2004spatial-dirac}
Andrew~M Childs and Jeffrey Goldstone.
\newblock Spatial search and the dirac equation.
\newblock {\em Physical Review A}, 70(4):042312, 2004.

\bibitem{childs2014spatial-crystal}
Andrew~M. Childs and Yimin Ge.
\newblock Spatial search by continuous-time quantum walks on crystal lattices.
\newblock {\em Phys. Rev. A}, 89:052337, May 2014.

\bibitem{Note4}
The cost of reflecting on this state can be quantified as $2\protect \mathcal
  {S}$. However, we shall omit constant factors for simplicity.

\bibitem{Note5}
The results of \cite {berry2017exponential} on simulating continuous query
  models can be used to quantify the cost of implementing time-evolution of
  $H_{search}$ in terms of discrete queries to the Grover oracle as well as the
  cost of simulating $H$ and the error of the simulation.

\bibitem{brassard2002quantum}
Gilles Brassard, Peter Hoyer, Michele Mosca, and Alain Tapp.
\newblock Quantum amplitude amplification and estimation.
\newblock {\em Contemporary Mathematics}, 305:53--74, 2002.

\bibitem{ambainis2005coins}
Andris Ambainis, Julia Kempe, and Alexander Rivosh.
\newblock Coins make quantum walks faster.
\newblock In {\em Proceedings of the sixteenth annual ACM-SIAM symposium on
  Discrete algorithms}, pages 1099--1108. Society for Industrial and Applied
  Mathematics, 2005.

\bibitem{ambainis2007quantum}
Andris Ambainis.
\newblock Quantum walk algorithm for element distinctness.
\newblock {\em SIAM Journal on Computing}, 37(1):210--239, 2007.

\bibitem{Note6}
For non vertex-transitive graphs, the structure can be such that certain
  particular nodes can be found faster than $\protect \sqrt {n}$ time. For
  example, the central node on a star graph can be found in constant time.
  However, if any of the graph's nodes can be marked, the minimum time needed
  to find any node in a graph is lower bounded by $\protect \sqrt {n}$ \cite
  {farhi1998analog}.

\bibitem{magniez2011search}
Fr{\'e}d{\'e}ric Magniez, Ashwin Nayak, J{\'e}r{\'e}mie Roland, and Miklos
  Santha.
\newblock Search via quantum walk.
\newblock {\em SIAM Journal on Computing}, 40(1):142--164, 2011.

\bibitem{krovi2016quantum}
Hari Krovi, Fr{\'e}d{\'e}ric Magniez, Maris Ozols, and J{\'e}r{\'e}mie Roland.
\newblock Quantum walks can find a marked element on any graph.
\newblock {\em Algorithmica}, 74(2):851--907, 2016.

\bibitem{ambainis2019quadratic}
Andris Ambainis, Andr\'{a}s Gily\'{e}n, Stacey Jeffery, and Martins Kokainis.
\newblock Quadratic speedup for finding marked vertices by quantum walks.
\newblock {\em arXiv:1903.07493}, 2019.~To appear in the Proceedings of 52nd
  Annual Symposium on the Theory of Computing (STOC) 2020.

\bibitem{chakraborty2018finding}
Shantanav Chakraborty, Leonardo Novo, and J{\'e}r{\'e}mie Roland.
\newblock Finding a marked node on any graph by continuous time quantum walk.
\newblock {\em arXiv preprint arXiv:1807.05957}, 2018.

\bibitem{wong2016quantum2}
Thomas~G Wong.
\newblock Quantum walk search through potential barriers.
\newblock {\em Journal of Physics A: Mathematical and Theoretical},
  49(48):484002, 2016.

\bibitem{wong2016laplacian}
Thomas~G Wong, Lu{\'\i}s Tarrataca, and Nikolay Nahimov.
\newblock Laplacian versus adjacency matrix in quantum walk search.
\newblock {\em Quantum Information Processing}, 15(10):4029--4048, 2016.

\bibitem{wong2016engineering}
Thomas~G Wong and Pascal Philipp.
\newblock Engineering the success of quantum walk search using weighted graphs.
\newblock {\em Physical Review A}, 94(2):022304, 2016.

\bibitem{glos2018vertices}
Adam Glos, Aleksandra Krawiec, Ryszard Kukulski, and Zbigniew Pucha{\l}a.
\newblock Vertices cannot be hidden from quantum spatial search for almost all
  random graphs.
\newblock {\em Quantum Information Processing}, 17(4):81, 2018.

\bibitem{Note7}
This is possible independently of the position of the marked node $\protect
  \ket {w}$, for example, if the Hamiltonian is diagonalized by the Fourier
  transform.

\bibitem{Boettcher_evaluation_2017}
Stefan Boettcher and Shanshan Li.
\newblock Evaluation of spectral zeta-functions with the renormalization group.
\newblock {\em Journal of Physics A: Mathematical and Theoretical},
  50(12):125003, feb 2017.

\bibitem{Note8}
From the Perron-Frobenius theorem, $\Delta _\protect \ensuremath {\protect
  \mathcal {D}}$ can never zero for the adjacency matrix (or Laplacian) of a
  connected graph.

\bibitem{Note9}
Since we are assuming $D$ eigenstates with eigenvalue $1$, $\protect \ket
  {v_n}$ can be any state in the subspace $\protect \ensuremath {\protect
  \mathcal {D}}$.

\bibitem{childs2019trotter}
Andrew~M. Childs, Yuan Su, Minh~C. Tran, Nathan Wiebe, and Shuchen Zhu.
\newblock A theory of trotter error.
\newblock {\em arXiv:1912.08854}, 2019.

\bibitem{glos2019impact}
Adam Glos and Tomasz Januszek.
\newblock Impact of global and local interaction on quantum spatial search on
  chimera graph.
\newblock {\em International Journal of Quantum Information}, 17(05):1950040,
  2019.

\bibitem{foulger2014quantum}
Iain Foulger, Sven Gnutzmann, and Gregor Tanner.
\newblock Quantum search on graphene lattices.
\newblock {\em Physical review letters}, 112(7):070504, 2014.

\bibitem{szegedy2004quantum}
Mario Szegedy.
\newblock Quantum speed-up of markov chain based algorithms.
\newblock In {\em Proceedings. 45th Annual IEEE Symposium on Foundations of
  Computer Science, 2004.}, pages 32--41. IEEE, 2004.

\bibitem{lucas2014ising}
Andrew Lucas.
\newblock Ising formulations of many np problems.
\newblock {\em Frontiers in Physics}, 2:5, 2014.

\bibitem{callison2019finding}
Adam Callison, Nicholas Chancellor, Florian Mintert, and Viv Kendon.
\newblock Finding spin-glass ground states using quantum walks.
\newblock {\em arXiv preprint arXiv:1903.05003}, 2019.

\bibitem{berry2017exponential}
Dominic~W Berry, Andrew~M Childs, Richard Cleve, Robin Kothari, and Rolando~D
  Somma.
\newblock Exponential improvement in precision for simulating sparse
  hamiltonians.
\newblock In {\em Forum of Mathematics, Sigma}, volume~5. Cambridge University
  Press, 2017.

\end{thebibliography}
\end{document}